%% file: 0_main_preprint.tex
\documentclass[%
 aip,
 amsmath,amssymb,
reprint,%
floatfix
]{revtex4-1}

\usepackage{graphicx}
\usepackage{dcolumn}
\usepackage{bm}
\usepackage{nicefrac}
\usepackage{mathtools}

\usepackage[utf8]{inputenc}
\usepackage[T1]{fontenc}
\usepackage{mathptmx}
\usepackage{etoolbox}
\usepackage[dvipsnames]{xcolor}
\usepackage{titlesec}
\usepackage{xr}
\makeatletter
\def\@email#1#2{%
 \endgroup
 \patchcmd{\titleblock@produce}
  {\frontmatter@RRAPformat}
  {\frontmatter@RRAPformat{\produce@RRAP{*#1\href{mailto:#2}{#2}}}\frontmatter@RRAPformat}
  {}{}
}%
\makeatother

\begin{document}
\title{
Topological analysis reveals multiple pathways in molecular dynamics
}

\author{Luca Donati}
\affiliation{Freie Universit\"at Berlin, Department of Mathematics and Computer Science, Arnimallee 22, D-14195 Berlin}
\affiliation{Zuse Institute Berlin, Takustr. 7, D-14195 Berlin, Germany}
\author{Surahit Chewle}
\affiliation{Zuse Institute Berlin, Takustr. 7, D-14195 Berlin, Germany}
\author{Dominik St. Pierre}
\affiliation{Freie Universit\"at Berlin, Department of Biology, Chemistry and Pharmacy, Arnimallee 22, D-14195 Berlin}
\affiliation{Zuse Institute Berlin, Takustr. 7, D-14195 Berlin, Germany}
\author{Vijay Natarajan}
\affiliation{Indian Institute of Science, Bangalore, India}
\affiliation{Zuse Institute Berlin, Takustr. 7, D-14195 Berlin, Germany}
\author{Marcus Weber}
\email{donati@zib.de}
\affiliation{Zuse Institute Berlin, Takustr. 7, D-14195 Berlin, Germany}

\begin{abstract}
Molecular Dynamics simulations are essential tools for understanding the dynamic behavior of biomolecules, yet extracting meaningful molecular pathways from these simulations remains challenging due to the vast amount of generated data. In this work, we present Molecular Kinetics via Topology (MoKiTo), a novel approach that combines the ISOKANN algorithm to determine the reaction coordinate of a molecular system with a topological analysis inspired by the Mapper algorithm. Our strategy efficiently identifies and characterizes distinct molecular pathways, enabling the detection and visualization of critical conformational transitions and rare events. This method offers deeper insights into molecular mechanisms, facilitating the design of targeted interventions in drug discovery and protein engineering.
\end{abstract}

\maketitle

%
\input{1_content}
\section*{Data Availability Statement}
The software used for this study is openly available on GitHub at \url{https://github.com/donatiluca/MoKiTo}. The complete dataset, including original MD trajectories and torch arrays representing pairwise distance matrices, is archived on Zenodo (DOI: \url{10.5281/zenodo.14229803}) and securely stored on the Zuse Institute Berlin server, and can be made available upon reasonable request.
\section*{Supporting Information}
\begin{itemize}
    \item Additional information regarding the implementation of the algorithm used to train the FNN within the ISOKANN method.
    \item Additional Figs.~S1, S2, S3, S4 showing the convergence of the $\chi$-function for each example presented in the manuscript, are available in the SI appendix. 
\end{itemize}

\begin{acknowledgments}
This research has been partially funded by the Deutsche Forschungsgemeinschaft (DFG, German Research Foundation) Cluster of Excellence MATH+, project AA1-15: ``Math-powered drug-design'', additionally it has been partially funded by the Bundesministerium für Bildung und Forschung (BMBF, Federal Ministry of Education and Research) within the project ``CCMAI -- Computermodellierung und k\"unstliche Intelligenz zur Aufkl\"arung von physiologischer und pathologischer Rezeptorfunktion''. We thank the DFG and the BMBF for their support. VN was supported by Berlin MATH+ under the Visiting Scholar program and by a research stay from the Alexander von Humboldt Foundation.
\end{acknowledgments}

\end{document}

%% file: 1_content.tex
Molecular Dynamics (MD) simulations are an indispensable tool for studying the dynamic behavior and properties of molecules and materials at the atomic level that are difficult or impossible to study experimentally \cite{karplus2002molecular, frenkel2001understanding}.
For example, MD simulations are employed to investigate key processes in biology such as protein folding or receptor-ligand binding, but also to model battery materials to improve energy storage technologies.

Despite their strength, MD simulations present several challenges: how to properly set up a simulation, the need for vast computational resources to simulate large systems as proteins, the long computational time required to complete a simulation.
Additionally, MD approach struggles with capturing chemically relevant events, often referred to as ''rare events'', such as slow conformational changes in proteins due to high energy barriers \cite{allen2017computer, laio2002escaping}.

\begin{figure}
    \centering
    \includegraphics[width=1\linewidth]{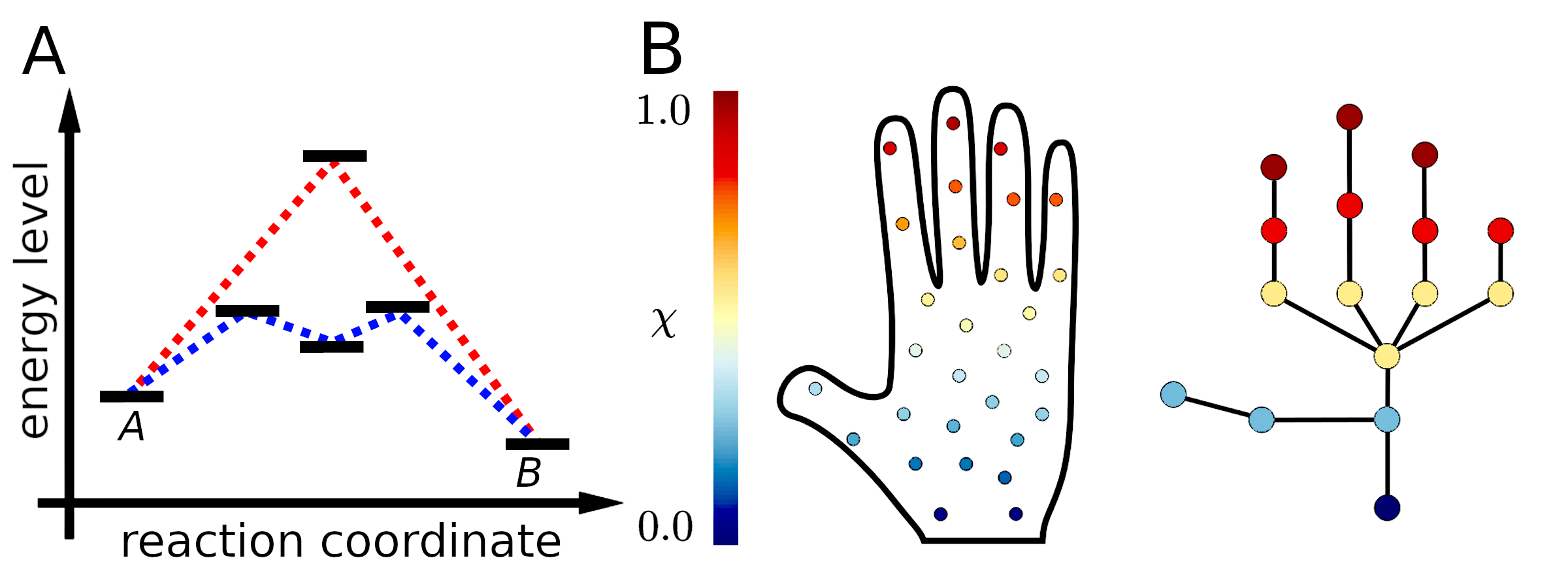}
    \caption{
    (A) Schematic representation of an energy diagram. 
    $A$ and $B$ represent two macro-states, while the intermediate states represent transition states. The dashed lines represent different pathways that the system can take to switch from macro-state $A$ to $B$ and vice versa.
    (B) Topological analysis of a ``hand'' using an ordering parameter. This is a re-elaboration of the example in Ref.~\cite{Lum2013}.
    }
    \label{fig:fig1}
\end{figure}

Once the simulations are computed, the analysis of MD simulations is a tedious task, in particular for long simulations of large systems.
Molecules are indeed high-dimensional systems, and in order to properly interpret the massive amounts of data generated by MD simulations, it is necessary to reduce dimensionality and find a human-readable representation.
%
%
%
The typical approach is to determine the macro-states (or metastable states) of the system and one or more reaction coordinates that describe the transitions between macro-states.
More precisely, given the state space $\Gamma$ of all conformational states, the macro-states are subsets of $\Gamma$ where the system can reside for a considerable amount of time and the reaction coordinate is a function $s:\Gamma\rightarrow [0,1]$ which assigns a real value from the interval $[0,1]$ to every conformational state $x \in \Gamma$.
For example, assuming the existence of two macro-states $A$ and $B$ and one reaction coordinate $s$, conformational states $x$ such that $s(x)\approx 0$ or $s(x)\approx 1$ compose the macro-states $A$ or $B$ respectively, while conformational states such that $s(x) \approx 0.5$ are considered transition states \cite{dellago1998efficient}.
Once the macro-states and the reaction coordinates have been identified, a diagram that represents the energy levels of the macro-states can be drawn as in Fig.~\ref{fig:fig1}-(A). 

This is an elegant and comprehensive way to represent molecular kinetics. 
The energy levels indicate which states are more stable, meaning they are more frequently visited during a simulation, and which pathways are more likely to occur.
However, the choice and the use of reaction coordinates can lead to an incomplete or inaccurate description of the system’s dynamics.
Important degrees of freedom may not be included in the chosen reaction coordinates; moreover, inappropriate dimensionality reduction violates Markovianity, an important assumption in MD simulations, where the system's future behavior depends only on its current state and not on its history.
These inaccuracies can lead to erroneous interpretations of the system behavior. 
Therefore, a systematic method for selecting optimal reaction coordinates that retain essential dynamic features and preserve Markovianity is required to ensure accurate simulation interpretations.

In this work, we propose a new strategy, called Molecular Kinetics via Topology (MoKiTo), to reproduce energy diagrams derived from MD simulations as in Fig.~\ref{fig:fig1}-(A) and Molecular Kinetics Maps (MKMs), i.e. graphical representations that highlight macro-states, transition states and pathways in between them.
The distinct feature of our approach is to use, as reaction coordinates, the ``membership functions'' $\chi_i:\Gamma \rightarrow [0,1]$.
The membership functions, from now on $\chi$-functions, were initially introduced in Refs.~\cite{Deuflhard2004,Weber2006thesis}, where they are the result of the numerical algorithm PCCA+, which operates a linear transformation to the eigenfunctions of the Koopman operator underlying the dynamics of the system.
Aside from technical details, which are clarified in the theory section, the significance of the $\chi$-functions is to describe the probability that a system's conformational state belongs or does not belong to a given macro-state.
A fundamental property of $\chi$-functions is that the projection of multidimensional dynamics onto them preserves Markovianity \cite{Weber2006thesis} and the dominant implied timescales of the Koopman operator projected onto $\chi$ are the same as those of the Koopman operator defined in full state space.
It follows, that $\chi$-functions are the optimal reaction coordinates for representing a high-dimensional system in a low-dimensional space.

Here, to construct the $\chi$-functions, we use the ISOKANN algorithm \cite{Rabben2020,Sikorski2024, Donati2024}, which through an iterative procedure and MD simulations as input data, trains an artificial neural network until it converges to a $\chi$-function.
The principal advantage of ISOKANN is that it is able to overcome the curse of dimensionality and estimate the $\chi$-function from a finite number of conformational states representing the state space.
On the other hand, ISOKANN assumes the existence of only two macro-states $A$ and $B$, and finds a unique membership function $\chi:=\chi_A = 1- \chi_B$.
The gradient of the $\chi$-function thus  makes it possible to determine the direction along which the slowest dynamics from $A$ to $B$ occurs \cite{weber2017fuzzy}, but it cannot determine alternative pathways from $A$ to $B$.

Here, to find the secondary pathways, we propose a new clustering technique inspired by the Mapper algorithm \cite{Gurjeet2007}.
The Mapper algorithm is a topological data analysis technique used to extract and visualize the underlying structure of high-dimensional data as a graph. 
Mapper begins by applying a filter function to project the data into a lower-dimensional space, and then partitions the data into overlapping clusters. 
These clusters are connected based on shared data points, forming a graph that captures the data's geometric and topological properties. 
As an example, already given in Ref.~\cite{Lum2013}, we show in Fig.~\ref{fig:fig1}-(B) how points sampled from a human hand can be clustered and connected in a graph using a filter function that assigns values from 0 to 1, from the wrist to the end of the middle finger.
In our context, the filter function is the $\chi$-function, which is used for a first partial clustering based on the macroscopic properties of the system.

We applied MoKiTo to several molecular systems to characterize the state space of the system and gain deeper insights into its dynamics. 
Of particular interest is the fourth example, where we studied the villin headpiece subdomain \cite{vardar2002villin,tang2006multistate,Lei2007,lindorff2011fast}, revealing both dominant and minor folding pathways.
This detailed mapping of the folding landscape not only highlights the complexity of protein folding, but also emphasizes the importance of alternative routes that, despite being less frequent, contribute significantly to the overall dynamics of the protein.
%


%
%
\section*{Background theory}
\label{sec2}
\subsection*{The Koopman operator}
We consider a molecular system defined in state space $\Gamma$ and denote by $x_t \in \Gamma$ a specific state at time $t$. 
We assume that the system is in equilibrium with a thermal reservoir at a constant temperature $T$ and that the dynamics is Markovian, ergodic, reversible and samples a unique stationary density $\pi(x)$.
We are especially interested in the time evolution of observable functions $f(x):\Gamma \rightarrow \mathbb{R}$, i.e. functions of the state of the molecule that describe some physical property of the system.
For this purpose, we introduce the operator $\mathcal{Q}$, also known in literature as backward Kolmogorov operator, that defines the partial differential equation
\begin{eqnarray}
  \frac{\partial f_t(x)}{\partial t} =
  \mathcal{Q} f_t(x) \, .
  \label{eq:Q}
\end{eqnarray}
The solution of eq.~\ref{eq:Q} is formally written as
\begin{eqnarray}
    f_{t+\tau}(x) 
    &=& 
    \exp \left(\mathcal{Q}\,\tau \right)f_t(x) \\
    &=&
    \mathcal{K}_{\tau}f_t(x) 
    \, .
    \label{eq:Koopman}
\end{eqnarray}
where we introduced the Koopman operator $\mathcal{K}_{\tau}$, defined in function space $L^{\infty}=\lbrace f: \Vert f \Vert_{\infty} < \infty \rbrace$, that propagates observable functions in time.
Eq.~\ref{eq:Koopman} can be also rewritten as a conditional expectation
\begin{eqnarray}
    f_{t+\tau}(x) 
    &=& 
    \mathbb{E}\left[ f_t(x_{t+\tau})\vert x_t = x\right]
    \, ,
    \label{eq:Koopman2}
\end{eqnarray}
where $f_t$ is applied to the states $x_{t+\tau}$ at time $t+\tau$, given an initial state $x$.
In most cases, we lack an analytical representation of the Koopman operator, however, given the ergodic property, the expectation form of the Koopman operator defined in eq.~\ref{eq:Koopman2} can be approximated as
\begin{eqnarray}
    f_{t+\tau}(x) 
    &\approx& 
    \frac{1}{M} \sum_{m=1}^M f_t(x_{t+\tau,m}\vert x_t = x)
    \, ,
    \label{eq:Koopman3}
\end{eqnarray}
where $x_{t+\tau,m}$ are the final states of $M$ trajectories of length $\tau$ starting at $x_t=x$, solutions of the equations of motion associated to the Koopman operator.
\subsection*{The $\chi$-function}
Let $\lambda_{\tau,i}$ and $\psi_i$ the eigenvalues and eigenfunctions of the Koopman operator that satisfy the eigenproblem
\begin{eqnarray}
    \mathcal{K}_{\tau} \psi_i = \lambda_{\tau,i} \psi_i\, .
\end{eqnarray}
Under the initial assumptions, the first eigenfunction is a constant function $\psi_0=1$ associated to the non-degenerate $\tau$-dependent eigenvalue $\lambda_{\tau,0}=1$.
All the other eigenfunctions have positive and negative values and are associated to sorted eigenvalues $1>\lambda_{\tau,1} \geq \lambda_{\tau,2} \geq \dots > 0$.

Let us now assume that the system is characterized by metastability, i.e. that there are $n_c$ macro-states where the system can remain for extended periods before changing macro-state.
Then, the first $n_c$ dominant eigenfunctions $\psi_i$ of the Koopman operator lie onto a $(n_c-1)$-simplex whose vertices represent the macro-states whereas the edges represent the transition paths between macro-states. 
By means of a linear transformation, it is possible to transform the simplex into a standard simplex, i.e. a simplex whose vertices are unit vectors.
Accordingly, the set of dominant eigenfunctions is transformed into a set of membership functions $\chi=\left( \chi_1,\chi_2,\dots,\chi_{n_c}\right)^\top$, with $\chi_i:\Gamma \rightarrow [0,1]$, $\forall i=1,2,\dots,n_c$, such that $\sum_i \chi_i = 1$.
In this way, the $\chi$-functions indicate the degree of membership of a specific macrostate.
For the case $n_c=2$, i.e. a bimetastable system, it is possible to derive the analytical solution \cite{Deuflhard2004}
\begin{eqnarray}
\begin{dcases}
    \chi_0(x) = \frac{\psi_1 - \min_x \psi_1(x)}{\max_x \psi_1(x) - \min_x \psi_1(x)}\, , \\
    \chi_1(x) = 
    1 - \chi_0(x)\, .
\end{dcases}
\label{eq:chi01}
\end{eqnarray}
Instead, for $n_c>2$, it is necessary to solve an optimization problem by means of the PCCA+ algorithm \cite{Deuflhard2004,Weber2006thesis}.

From here on, we will assume the case $n_c=2$, and since the functions $\chi_0$ and $\chi_1$ are complementary, we will refer to one of them simply by the notation $\chi$.

%
\section*{Methods}
\label{sec3}
\begin{figure*}[bt!]
    \includegraphics[width=1\textwidth]{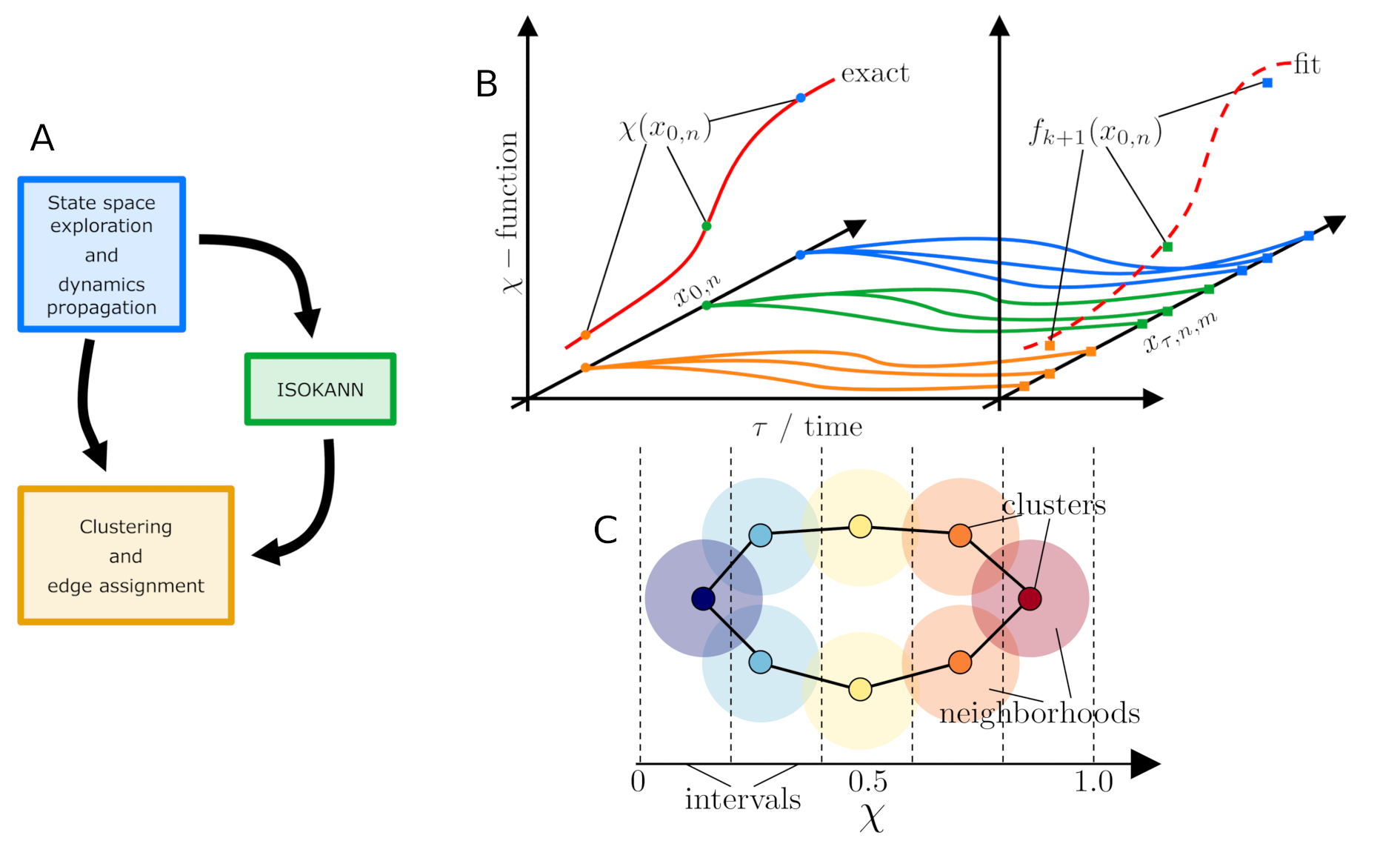}
    Description of the methods.
    \caption{(A) MoKiTo workflow diagram. Constructing the MKM using a three-stage procedure.
    (B) In the case of unknown $\chi$-function, it is necessary first to propagate short trajectories starting from $x_{0,n}$, then to apply an arbitrary function $f_k$ and estimate the average for each state $x_{0,n}$, and then to apply the shift-scale function $S$ as in eq.~\ref{eq:Koopman4}.
    ISOKANN scheme. Given the exact  $\chi$-function, it is possible to calculate $\chi(x_{0,n})$ as on the left side.
    The states thus found are then connected via a fit function obtained by a FNN.
    (C) Clustering and edge assignment scheme. The states representing the state space are first subdivided into intervals according to the $\chi$ function. Then, the states of the same interval are clustered by CNN clustering algorithm and edges are found by overlapping the neighborhoods.
    }
    \label{fig:fig2}
\end{figure*}
In this section, we describe the three-stage procedure MoKiTo, as outlined in Fig.~\ref{fig:fig2}-(A), for constructing the MKM:
\begin{itemize}
    \item The first stage focuses on exploration of the state space and propagation of dynamics by means of MD simulations.
    \item The second stage involves construction of the $\chi$ function by means of the ISOKANN algorithm.
    \item The third stage uses a clustering algorithm to cluster the MD data filtered by the $\chi$-function.
    Then the edges of the graph are assigned according to the overlap of neighborhoods.
    The outcome of this procedure is a MKM.
\end{itemize}

\noindent \textbf{State space exploration and dynamics propagation.}
The objective of this stage is to identify $N$ states representative of state space $\Gamma$.
This can be achieved in different ways depending on the system under investigation.
As we will show in our numerical experiments, for systems characterized by low energy barriers, e.g., short chains of amino acids, conventional Molecular Dynamics or Monte Carlo simulations are the most convenient solution.
Instead, for bigger systems such as proteins, we recommend the use of enhanced techniques, such as
Simulated Tempered MD (STMD) simulations \cite{Marinari_1992},
replica exchange MD \cite{Sugita2000},
umbrella sampling \cite{Souaille2001}, 
or metadynamics \cite{Huber1994, Laio2002,Barducci2008}.
Here, in the fourth example of chicken villin headpiece protein, we opted for STMD simulations, where the problem of getting stuck in local minima
is overcome by dynamically adjusting the temperature during the simulation.
STMD simulations are advantageous over the other proposed methods because they do not require a set of collective variables to be chosen a priori. 
However, states generated by STMD simulations require further minimization and equilibration to ensure they represent the canonical ensemble. 

Once a set of $N$ representative states has been built, it is necessary to propagate the dynamics.
For this purpose, we perform $M$ conventional MD simulations of length $\tau$ for each representative state.
We would like to point out that, in this step, the simulations must represent the correct dynamics of the system under investigation, as they represent the action of the Koopman operator.
Thus, enhanced techniques that alter the potential energy function or the system temperature cannot be applied. 

For the later stages of the MKM construction, it is useful to organize the data using multidimensional arrays.
Here in the text, and in the attached Python scripts, we will use the following notation: 
\begin{eqnarray}
X_0=\lbrace x_{0,1}, \, x_{0,2},\, ..., \, x_{0,N} \rbrace \, 
\end{eqnarray} 
is the array of initial states with shape $(N,N_D)$;
\begin{eqnarray}
X_{\tau}=
\lbrace &x_{\tau,1,1},\, x_{\tau,1,2},\, ..., \, x_{\tau,1,M}; \\ &x_{\tau,2,1},\, x_{\tau,2,2},\, ..., \, x_{\tau,2,M}; \\ 
& ...
\\
& x_{\tau,N,1}, \, x_{\tau,N,2}, ..., \, x_{\tau,N,M} \rbrace
\end{eqnarray} is the array of final states of short trajectories with shape $(N,M,N_D)$.
$N_D$ indicates the number of dimensions used to represent each state of the system.
In the following examples, it is the number of Cartesian coordinates, the number of pairwise distances or the number of internal coordinates.\\

\noindent \textbf{ISOKANN.}
In the second stage of our procedure, we determine the $\chi$-function by means of the ISOKANN algorithm \cite{Rabben2020}, a modification of the Von-Mises algorithm \cite{Mises1929}.
In the original Von-Mises algorithm, given a square matrix $A$ and an initial arbitrary vector $v_0$, the iterative calculation
\begin{eqnarray}
    f_{k+1} &=& \frac{
    A v_k}{\lVert A v_k \rVert} \, ,
    \label{eq:Von_Mises}
\end{eqnarray}
converges to the dominant eigenvector of the matrix $A$ as $k\rightarrow \infty$.
However, in the case of the Koopman operator, or its matrix representation, the Von-Mises algorithm always converges to a constant vector, corresponding to the first eigenfunction $\psi_0=1$.
Thus, this algorithm is of no particular utility when applied to the Koopman operator.

ISOKANN, modifies the Von-Mises algorithm as
\begin{eqnarray}
    f_{k+1} &=& 
    S\mathcal{K}_{\tau} f_k\, ,
    \label{eq:Isokann}
\end{eqnarray}
where the function $S$ is the linear transformation
\begin{eqnarray}
    S\mathcal{K}_{\tau} f_k = \frac{\mathcal{K}_{\tau} f_k - \min\left(\mathcal{K}_{\tau} f_k\right)}{\max\left(\mathcal{K}_{\tau} f_k\right) - \min\left(\mathcal{K}_{\tau} f_k\right)} \, .
    \label{eq:S}
\end{eqnarray}
The function $S$, based on the definition of $\chi$-function for a bimetastable system in eq.~\ref{eq:chi01}, is known as shift-scale function.
It prevents the convergence of the initial function to the dominant eigenvector, guarantees that $f_k:\Gamma \rightarrow [0,1]$ and forces the convergence to the $\chi$-function.
Since we do not know an analytical expression of $\mathcal{K}_{\tau}$, nor a matrix representation of it, we exploit eq.~\ref{eq:Koopman3} and the trajectories generated in the previous stage.
Thus, the ISOKANN algorithm applies to an initial state $x_{0,n} \in X_0$ as
\begin{eqnarray}
    f_{k+1}(x_{0,n}) 
    &=& 
    S
    \frac{1}{M} \sum_{m=1}^M f_k(x_{\tau,n,m}\vert x_0 = x_{0,n})
    \, ,
    \label{eq:Koopman4}
\end{eqnarray}
where $x_{\tau,n,m}\in X_{\tau}$ is the final state of the $m$th trajectory started in $x_{0,n}$.
As $k\rightarrow \infty$, we obtain
\begin{eqnarray}
    \lim_{k\rightarrow \infty } f_{k+1}(x_{0,n}) &=& \chi(x_{0,n}).
\end{eqnarray}

In this formulation, we are assuming to have an analytical expression for $f_k(x)$, however the application of the eq.~\ref{eq:Koopman4} provides $f_{k+1}(x_{0,n})$ as scalar value.
We do not have an analytical expression for $f_{k+1}$, so how can we apply the equation to the next step?
To overcome this, we look for the analytical function that best fits the $N$ values $f_{k+1}(x_{0,n})$.
For regression, a wide range of options is
available: for low-dimensional systems, spline functions and radial basis functions may be better choices, given the low number of parameters to be trained, but for high-dimensional systems, the use of Feedforward Neural Networks (FNNs) is recommended.
In the examples presented in this manuscript, we employed a FNN, whose training procedure is described in the SI Appendix.
Fig.~\ref{fig:fig2}-(B) summarizes the ISOKANN procedure.\\

\noindent \textbf{Clustering and edge assignment.}
The last stage is devoted to the construction of the MKM.
We subdivide the $\chi$-function into $L$ disjoint intervals containing states with similar $\chi$-value.
The number of intervals into which the $\chi$-function is subdivided is arbitrary, but ideally there should be a sufficient number of states representing similar macroscopic behaviour, i.e. macro-states and transition states.
This partial clustering based on the $\chi$-function is useful to reduce  the complexity of the data while preserving important properties such as Markovianity \cite{Weber2018}. 

Once the intervals of the $\chi$-function have been defined, we cluster states in state space $\Gamma$.
Several algorithms could be used.
Here,  we chose the Common Nearest Neighbor (CNN) clustering algorithm \cite{Keller2010}, an unsupervised clustering algorithm that uses local density information to identify clusters of data points without prior knowledge of the number of clusters.
CNN clustering tends to work well with non-linearly separable data and has already been shown to be suitable for the study of molecular systems \cite{Lemke2016, Lemke2019}.
The key assumption of this algorithm is that two states are more likely to belong to the same cluster if they share a significant number of neighbors.
The algorithm is then controlled by two key parameters: the radius of the neighborhood $\varepsilon$ and the number of nearest neighbors $\theta$. 
Two states are considered neighbors if they are less than a distance $\varepsilon$ apart from each other:
\begin{eqnarray}
    x_i \ \mathrm{and} \ x_j \ \mathrm{are} \ 
    \mathrm{neighbors} \ \mathrm{if} \ 
    |x_i - x_j| < \varepsilon \, .
\end{eqnarray}
Then, for every pair of states, the intersection of their respective nearest neighbor sets is determined: if $x_i$ and $x_j$ share at least $\theta$ neighbors, they belong to the same cluster.
The algorithm determines $K$ clusters $\Omega_1,\, \Omega_2, \, ... , \, \Omega_K$, where $K$ is an output parameter, however, like many density-based methods, its performance is sensitive to the choice of $\theta$ and $\varepsilon$.
Following the indications in Refs.~\cite{Keller2010, Lemke2016}, the parameters should be chosen on the basis of the histogram of pairwise Root Mean Square Distances (RMSDs) between the states in the dataset: $\varepsilon$ should be set to a value slightly smaller than the first maximum of this histogram, while $\theta$ is varied until adequated sized clusters are found.

The size of the clusters is proportional to the Boltzmann weight in the canonical ensemble, so it is possible to derive an estimate of the cluster energy level from the well-known formula
\begin{eqnarray}
    E_{\Omega_i} = -\frac{1}{\beta} \log \pi_{\Omega_i} \, ,
    \label{eq:energy_cluster}
\end{eqnarray}
where $\pi_{\Omega_i}$ is the normalized size of the $i$th  cluster and $\beta=\nicefrac{1}{k_B T}$, with temperature $T$ and Boltzmann constant $k_B$.

The last step in the construction of the MKM is the assignment of edges, i.e. finding the connections between pairs of clusters $\Omega_i$ and $\Omega_j$ $\forall i,j=\,1,\,2,\,...,\,K$.
A transition in an infinitesimal span of time can only occur between similar conformations, both macroscopically and microscopically.
Thus, to determine the edges, we make the following assumptions:
\begin{itemize}
    \item  No transition occurs between clusters belonging to the same interval.
    \item An infinitesimal transition can only occur between clusters belonging to consecutive intervals that have common states in their neighborhood.
\end{itemize}
The first assumption is motivated by the observation that two clusters within the same interval, i.e., with similar $\chi$ values, exhibit macroscopic similarity but microscopic differences, likely due to the presence of an energy barrier.
Such transitions are rare, and we therefore choose to neglect them.
In contrast, the gradient of the $\chi$ function determines the direction of the dominant process. 
Consequently, a transition between clusters in consecutive intervals corresponds to a transition between states with similar macroscopic properties.
Additionally, we require that the two clusters share states in their neighborhood, ensuring that the transition occurs between states with comparable structural properties.
This procedure begins by aligning the states within a cluster to minimize the RMSD and computing the average structure of the cluster.
Next, the neighborhood of the cluster, whose size is determined by a threshold $r_n$, is identified by calculating the RMSD between the average structure and all the states in the $X_0$ dataset, which contains the representative states of the state space.

The procedure to determine the clusters and the edges of the MKM is schematized in Fig.~\ref{fig:fig2}-(C).

%
\section*{Results}
\label{sec4}
%
%
\subsection*{Two-dimensional system}
\begin{figure*}[bt!]
    \centering
    \includegraphics[width=1\textwidth]{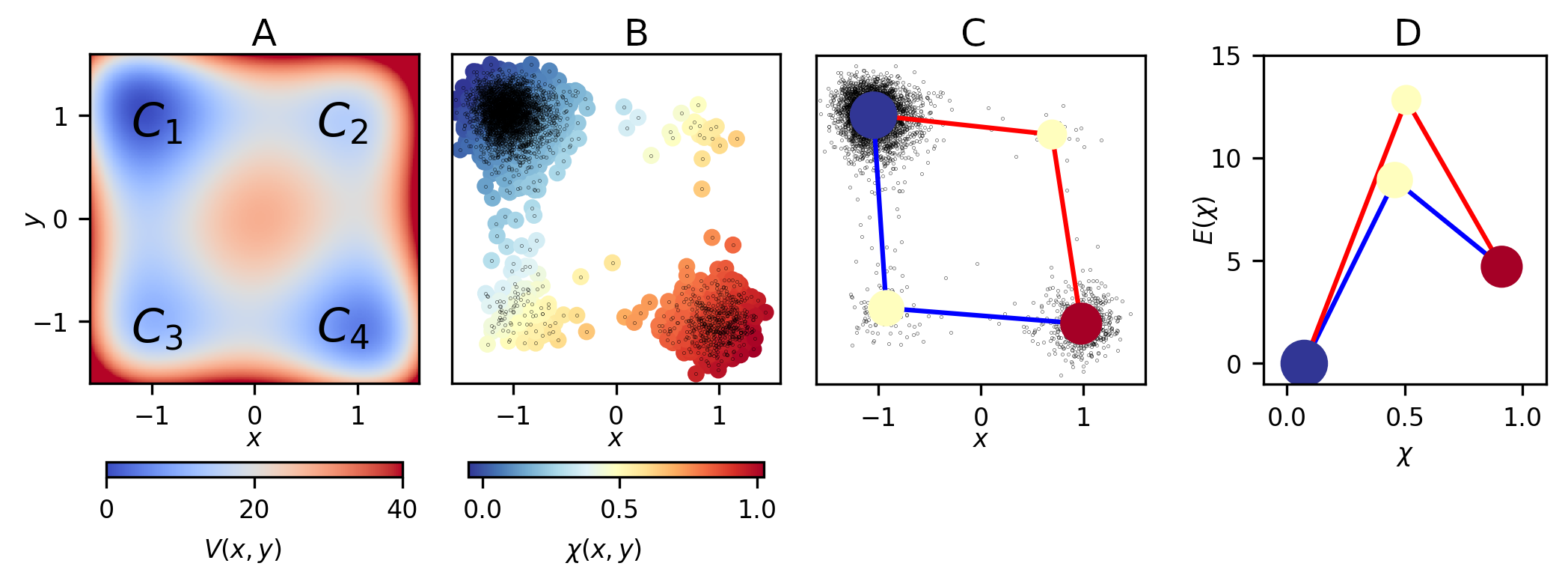}
    \caption{
    Results of the two-dimensional system.
    (A) Potential energy function of the two-dimensional system; 
    (B) States of the two-dimensional system extracted from a trajectory and colored according to the membership function $\chi(x,y)$.
    (C) MKM of the two-dimensional system projected onto the Cartesian space. The black dots represent the initial states $X_0$;
    (D) Energy landscape of the two-dimensional system.
    }
    \label{fig:fig3}
\end{figure*}
As an illustrative example, we considered a two-dimensional system governed by overdamped Langevin dynamics and defined by the potential energy function
\begin{eqnarray}
    V(x,y)       = 10(x^2 - 1)^2 + 5xy + 10(y^2-1)^2 + 2.2x  \, ,
\end{eqnarray}
illustrated in Fig.~\ref{fig:fig3}-(A).
The potential is characterized by 4 local minima of different height: the deepest is corner $C_1=(-1,1)$, followed by $C_4=(1,-1)$, $C_3=(-1,-1)$ and $C_2=(1,1)$, in order from lowest to highest .
To give physical meaning to the problem, we assumed that the potential has energy units $\mathrm{kJ\, mol^{-1}}$, and that generates forces $-\nabla_x V$ and $-\nabla_y V$ on a fictitious particle of mass $m = 1\, \mathrm{amu}$ that moves on a flat surface.
We also assumed standard thermodynamic parameters: the temperature of the system was $T=300\, \mathrm{K}$ with molar Boltzmann constant  $k_B = 8.314\times10^{-3} \,\mathrm{kJ\, K^{-1}\, mol^{-1}}$.
This choice of the parameters makes sure metastability, indeed the system's thermal energy $\beta^{-1} = k_B T = 2.49 \, \mathrm{kJ\, mol^{-1}}$ is significantly smaller than the height of the barriers along $x$ and $y$.
The interaction of the particle with the environment is modeled via a friction coefficient $\gamma=1\,\mathrm{ps}^{-1}$ and a diffusion constant $D = \nicefrac{k_B T}{m \gamma} = 2.49\,\mathrm{nm^2\,ps^{-1}}$ in each direction.\\
%

%
\noindent \textbf{State space exploration and dynamics propagation.}
We solved the overdamped Langevin dynamics equations of motion
\begin{eqnarray}
    \begin{cases}
        dx_t = - \beta D \, \nabla_x V(x_t,y_t) \, dt + \sqrt{2D} \, dW_x \\
        dy_t = - \beta D \, \nabla_y V(x_t,y_t) \, dt + \sqrt{2D} \, dW_y \\
    \end{cases} \, ,
\end{eqnarray}
where $W_x$ and $W_y$ are two independent and uncorrelated Wiener processes, applying the Euler-Maruyama scheme \cite{Leimkuhler2015} with a timestep of $\Delta t = 0.001 \, \mathrm{ps}$.

First, we generated a sufficiently long trajectory of $1\times 10^7$ timesteps which covers the relevant regions of the potential.
Then, we extracted 4000 initial states equally spaced from the trajectory, i.e. one each 1000 timesteps, and carried out 10 short trajectories of 10 timesteps from each initial state.
%

%

As suggested in the methods section, we organized the data into two arrays: $X_{0}$ of shape (4000,2) containing the coordinates of the initial states, and $X_{\tau}$ of shape (4000,10,2) containing the coordinates of the final states of the short trajectories.\\

\noindent \textbf{ISOKANN.}
In order to construct the $\chi$-function, we applied ISOKANN.
For regression, we used an FNN with three layers and the the sigmoid function as activation function.
The FNN was implemented using torch \cite{NEURIPS2019_9015}, the input layer had 2 nodes, one for each coordinate of the system, the hidden layer had 128 nodes and the output layer had 1 node corresponding to the $\chi$-value.
The optimization of the FNN parameters was realized using the Stochastic Gradient Descent (SGD) algorithm \cite{Robbins1951ASA} and minimizing the mean squared error (squared $\ell^2$-norm).
At each ISOKANN iteration, we performed a training of 15 epochs, iterating over randomly generated batches of size 100, with an initial learning rate of 0.001.
This choice of hyperparameters is the result of a random search and leads to a convergence of the $\chi$-function in 26 iterations (SI Appendix, Fig.~S1).

The $\chi$-function, evaluated in each initial point $X_0$, is illustrated in Fig.~\ref{fig:fig3}-(B).
The corners $C_1$ and $C_4$, colored by blue ($\chi\approx 0$) and red ($\chi\approx 1$) respectively, are the two main macro-states, i.e. the regions of state space most visited by the particle.
Corners $C_3$ and $C_2$ are colored with a gradient of colors blue-yellow ($\chi\approx 0.5$), and can be interpreted as transition state states.\\

\noindent \textbf{MKM construction.}
To construct the MKM representing the macrostates and the main pathways of the dynamics, we proceeded in two steps: first we grouped the states by dividing the $\chi$-function into 3 regular intervals and finding for each interval 1703, 32 and 263 states $(x,y)$ respectively, then we applied the CNN clustering algorithm to group the states into smaller clusters having similar $\chi$-value.
For the CNN clustering algorithm, we chose as radius of the neighborhood $\varepsilon=1.0$ and as number of nearest neighbors $\theta=5$, finding $N_c = 4$ clusters.
To connect the clusters, we searched for areas of overlap between neighborhoods of clusters belonging to consecutive intervals, using as a threshold the Euclidean distance $r_n = 0.6$.
The MKM is shown in Fig.~\ref{fig:fig3}-(C).\\

\noindent \textbf{Observations.}
The potential energy function has four metastable states and by means of the $\chi$-function we identify three macroscopic regions, two metastable regions for corners $C_1$ and $C_4$ respectively, and one transition region that includes both corners $C_2$ and $C_3$.
There are two main pathways from corner $C_1$ to corner $C_4$ and vice versa.
In Fig.~\ref{fig:fig3}-(D), we show the energy diagram, i.e. the energy levels of the clusters as defined in eq.~\ref{eq:energy_cluster}, connected by the edges of the MKM.
This representation is useful for interpreting the weight of clusters and evaluating the energy cost of clusters.
Since the blue and red clusters are the largest, their energy levels are low, the yellow clusters, instead, are higher and correspond to transition states.
From this graph, we also deduce that the red path ($C_1-C_3-C_4$) is the most likely path, as the system needs less energy to visit corner $C_3$ than corner $C_2$.
%

%
%
\subsection*{33-Dichloroisobutene}
\begin{figure*}[bt!]
    \centering
    \includegraphics[width=1\textwidth]{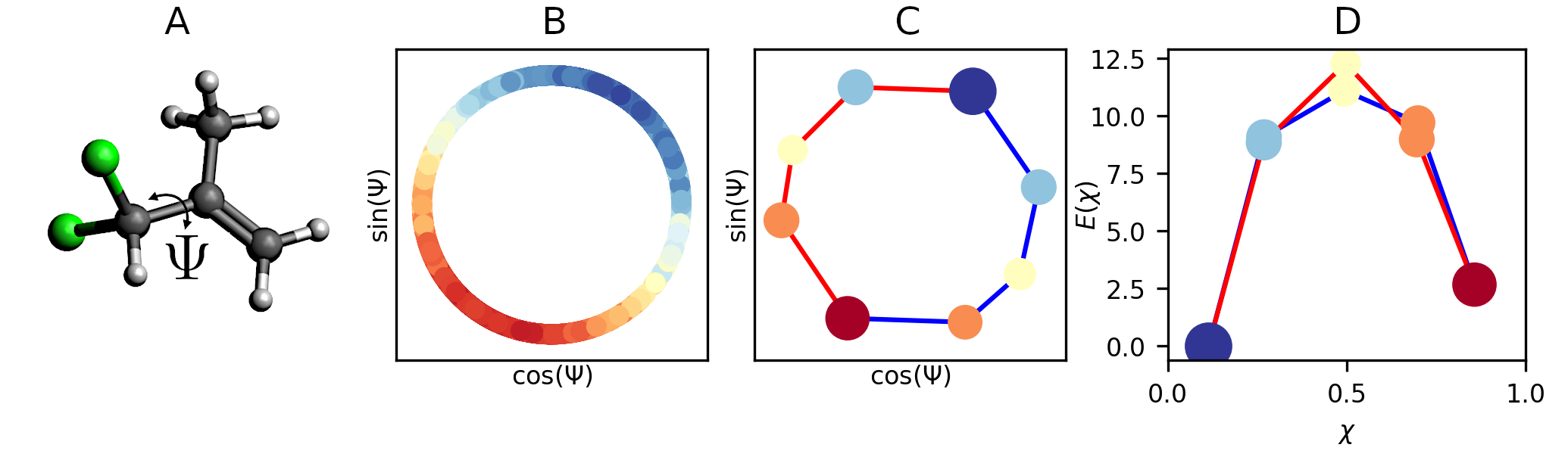}
    \caption{
    Results of 33-Dichloroisobutene molecule.
    (A) 33-Dichloroisobutene molecule; 
    (B) $\chi$-function projected onto the main torsion angle of the 33-Dichloroisobutene;
    (C) MKM projected onto the main torsion angle of the 33-Dichloroisobutene;
    (D) Energy landscape of the 33-Dichloroisobutene molecule.
    }
    \label{fig:fig4}
\end{figure*}
As the first molecular system example, we studied 3,3-Dichloroisobutene (C$_4$H$_6$Cl$_2$), a dichloro derivative of isobutene, represented in Fig.~\ref{fig:fig4}-(A).
The compound has 12 atoms, for a total of 36 dimensions, and the rotation around the torsion angle $\Psi$ (C$_3$-C$_2$-C$_4$-Cl$_1$) is known to be the slowest process of the system.
Thus, we used the torsion angle $\Psi$ as relevant coordinate to visualize the results.\\

\noindent \textbf{State space exploration and dynamics propagation.}
We performed MD simulations using the package OpenMM \cite{OpenMM2017} with the Generalized Amber Force Field \cite{Wang2004}.
To simulate an explicit solvent, we used the TIP3P-FB water model \cite{wang2014building} with a padding distance of 1.4 nm which generates a box of 682 water molecules for 33-Dichloroisobutene.
We assumed Langevin dynamics and applied the Langevin leapfrog integrator \cite{Izaguirre2010} with $\gamma=1.0\,\mathrm{ps}^{-1}$ as friction coefficient, and $\Delta t=2\, \mathrm{fs}$ as integrator timestep.
Non-bonded interactions between atoms, such as Coulomb forces and Lennard-Jones forces, were calculated by Particle-Mesh Ewald (PME) method \cite{Darden1993} and interactions between atoms more than 1 nm apart were truncated.
The lengths of all bonds involving a hydrogen atom have been constrained.
Before doing the first simulation, we brought the system to a local energy minimum, then we equilibrated the system with a 20 ps simulation to obtain a state belonging to the NVT ensemble, with temperature equal to $300\pm6\,\mathrm{K}$.

From a trajectory of $40\times10^6$ timesteps, corresponding to 80 ns, we extracted 4000 states $x_0=\lbrace r_0, v_0\rbrace$ (positions and velocities of each atom including the solvent) every 8 ps.
Then the states $x_0$ have been used as initial states for 10 new short trajectories of length 0.02 ps.\\

\noindent \textbf{ISOKANN.}
The procedure for constructing the $\chi$-function via ISOKANN was the same as in the previous example.
However, instead of providing the Cartesian coordinates of the atoms, we used the pairwise distances between all the atoms of the system (without the water).
This increases the number of dimensions of the $\chi$-function to $\nicefrac{12\cdot (12- 1)}{2}=66$, but ensures that $\chi$ is invariant with respect to translations and rotations.
As a model to approximate the $\chi$ function, we used a FNN with 4 layers (2, 66, 33, 1 nodes), however, as activation function we used Leaky ReLU which performs better than sigmoid in high-dimensionality regression tasks.
We set the initial learning rate for the SGD algorithm to 0.01 and applied a weight decay of 0.01 for regularization.
The learning process took place rather quickly and after 279 iterations, both the training loss and the validation loss were smaller than $10^{-3}$ (SI Appendix, Fig.~S2).
The $\chi$-function, evaluated in each initial point $x_0$, is reported in Fig.~\ref{fig:fig4}-(B).
For ease of reading, we have projected the $\chi$-function onto the unit circle, i.e. the values $\cos(\Psi)$ and $\sin(\Psi)$, where $\Psi$ is the torsion angle.
We clearly distinguish metastable states colored with red and blue, and the transition states colored with yellow.
The correlation between $\chi$-function and torsion angle $\Psi$ is equal to 0.9, confirming that the latter is a good choice as a relevant coordinate to describe the slowest process of the system.\\

\noindent \textbf{Clustering and edge assignment.}
To construct the MKM, we discretized the $\chi$-function into 5 equal intervals, then applied the CNN clustering algorithm.
Clustering was done by pre-computing the Root Means Square Distance (RMSD) matrix.
Then, after analyzing the distribution of RMSDs, we identified the CNN clustering parameters: $\varepsilon=0.09,\, 0.08,\,0.09,\, 0.07, \,0.09$, and $\theta=5$ (for each interval).
With this setting, we obtained 8 clusters, 2 for metastable states and 6 for transition states.
To find edges, we used as threshold for the neighborhoods $r_n=0.05$.
The MKM projected onto the unit circle of the angle $\Psi$ is shown in Fig.~\ref{fig:fig4}-(C).
\\

\noindent \textbf{Observations.} 
We observe two main clusters: the red one ($\Psi \approx \nicefrac{\pi}{4}$) corresponds to a conformational state where both chlorines are staggered with the methylene group; the blue one ($\Psi \approx \nicefrac{3\pi}{4}$) corresponds to a conformational state where both chlorines staggered with the methyl group.
The two metastable states are connected by two pathways that correspond to the rotations of the torsion angle $\Psi$.
The energy diagram, illustrated in Fig.~\ref{fig:fig4}-(D), also reveals the energy barrier of the paths.
Since the system is perfectly symmetrical with respect to the rotation of the $\Psi$ angle, the two paths overlap. 
In other words, no direction is preferred, and the system can rotate clockwise or counterclockwise with the same probability.
However, we observe that there is a higher barrier between the red cluster and subsequent orange clusters than between the blue cluster and subsequent light blue clusters.
Then the configurational states belonging to the red cluster are the most stable of the system.
%
\subsection*{Hexapeptide VGVAPG}
\begin{figure*}[bt!]
    \centering
    \includegraphics[width=1\textwidth]{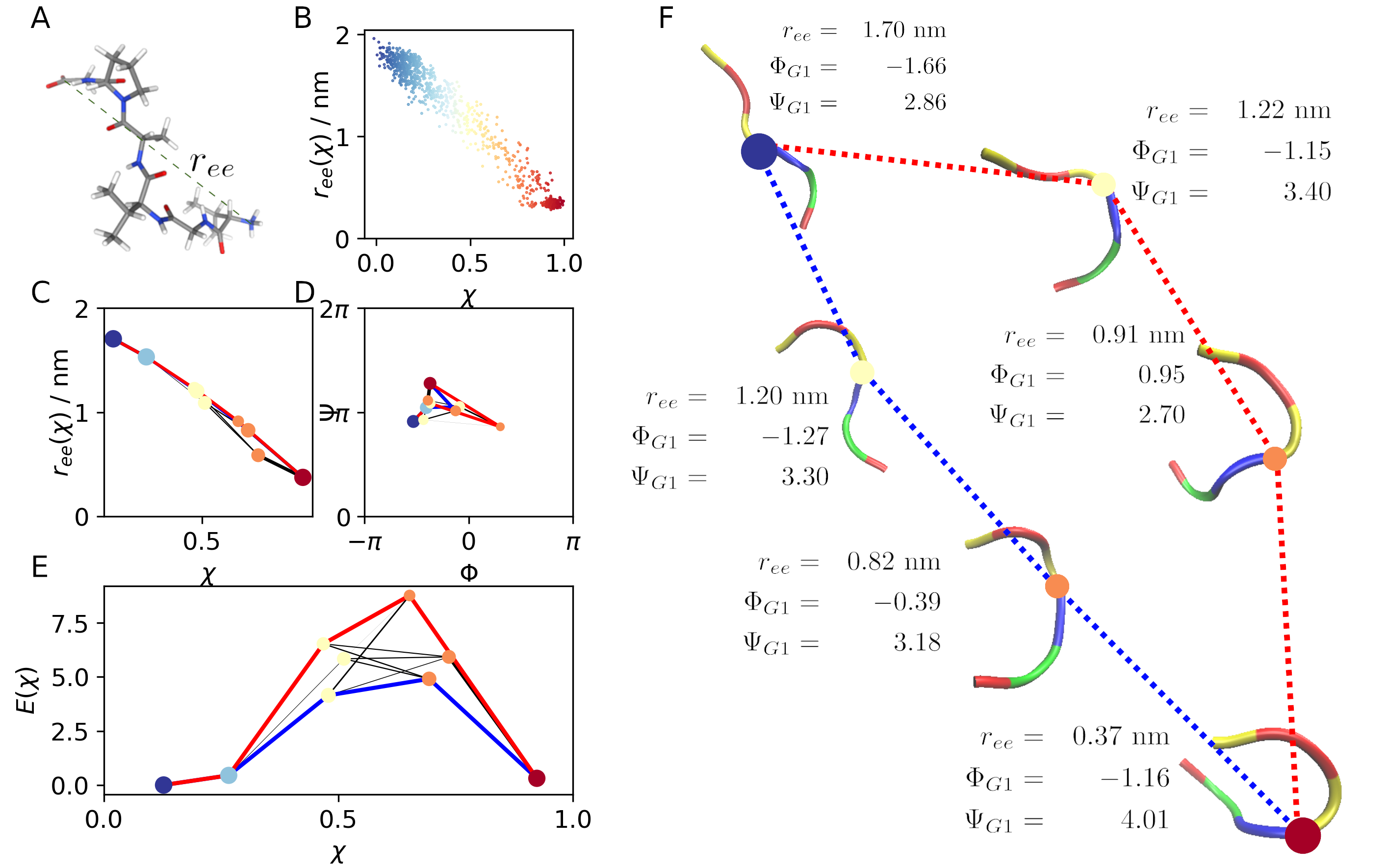}
    \caption{
    Results of VGVAPG hexapeptide.
    (A) VGVAPG molecule; 
    (B) $\chi$-function projected onto the end-to-end distance of the molecule.
    (C) MKM with representative structures  of the VGVAPG; 
    (D) MKM projected onto the end-to-end distance of the VGVAPG.
    (E) MKM projected onto the Ramachandran plot of the second residue (Glycine 1);
    (F) Energy landscape of VGVAPG with the most relevant clusters of the MKM. The colors of the molecular structure represent the residues: Val (yellow), Gly (red), Pro (blue) and Ala (green).
    }
    \label{fig:fig5}
\end{figure*}
The VGVAPG is an elastin-derived hexapeptide \cite{Floquet2004}, already used to test methods for MD simulations
 \cite{Donati2018, Barreto2021}.
The peptide has 73 atoms, corresponding to 219 dimensions in the Euclidean space.
As relevant coordinate, denoted by $r_{ee}$ in the figures, we used the Euclidean distance between the nitrogen atom of the N-terminus and the carboxyl-carbon of the C-terminus (Fig.~\ref{fig:fig5}-(A).\\

\noindent \textbf{State space exploration and dynamics propagation.}
The MD simulations were carried out with the same settings as for 33-Dichloroisobutene, but the water box was increased to 782 water molecules and the force field was the AMBER ff-14sb \citep{Maier2015}.
The length of the first simulation was $500\times 10^6$ timesteps, corresponding to $1\,\mu s$, from which, we extracted 1000 initial states for the short trajectories.
For each initial state, we produced 10 short trajectories of 1000 timesteps, corresponding to 2 ps.\\

\noindent \textbf{ISOKANN.}
We performed a random search to find the best hyper-parameters of the neural network and determined as optimal parameters, 1752 nodes in the hidden layer,  0.001 as initial learning rate and 0.005 as weight decay.
The ISOKANN algorithm was performed for 57 iterations, until the training and validation loss stabilized at approximately $4\times 10^{-4}$ (SI Appendix, Fig.~S3).
The $\chi$-function is plotted in Fig.~\ref{fig:fig5}-(B).
We observe a large metastable state, corresponding to $\chi\approx 0.0$ (blue), which includes configurations whose relevant coordinate $r_{ee}$ ranges from 0.7 to 2 nm; the transition states, with $\chi\approx 0.5$ (yellow), range from 0.35 to 1.7 nm and the metastable state corresponding to $\chi\approx 1.0$ (red), includes configurations whose relevant coordinate ranges from 0.25 to 0.35 nm.
The correlation between $\chi$ and $r_{ee}$ is 0.98. \\

\noindent \textbf{MKM construction.}
We divided the $\chi$-function into 5 equal intervals between 0 and 1.
Then, by analyzing the distribution of RMSDs, we determined the parameters for the CNN clustering: $\varepsilon=0.3, 0.25, 0.2, 0.17, 0.3$ and $\theta=5$ for each interval.
Thus, we found 1, 1, 3, 3 and 1 cluster for each interval respectively.
The MKM, obtained with $r_n=0.2$, is reported in Fig.~\ref{fig:fig5}, where we propose different representations: in Fig.~\ref{fig:fig5}-(C), we show the complete MKM projected onto the end-to-end distance of the peptide; in Fig.~\ref{fig:fig5}-(D) we show the MKM projected onto the Ramachandran plot of the first Glycine (G1) of the peptide; in Fig.~\ref{fig:fig5}-(E), we show the energy diagram; 
in Fig.~\ref{fig:fig5}-(F), we show the main representative structures of the backbone of the peptide (omitting the less relevant clusters of the MKM); 
\\

\noindent \textbf{Observations.}
The blue cluster ($\chi\approx 0$) comprises completely open structures with $r_{ee}> 1.5 \, \mathrm{nm}$, while the red cluster ($\chi\approx 1$) represents closed structures with $r_{ee} \approx 0.3 \, \mathrm{nm}$.
Since distance $r_{ee}$ is highly correlated with $\chi$, we cannot distinguish multiple paths from Fig.~\ref{fig:fig5}-(C).
Thus, to better characterize the dynamics and describe the opening-closing mechanism of the peptide, we analyzed the Ramachadran plot of each residue.
Here, in Fig.~\ref{fig:fig5}-(D), we report the Ramachadran plot of the second residue (the first Glycine in the chain VGVAPG), as it shows the most interesting dynamics.
First, we observe that most of the clusters, in particular the clusters belonging to the blue path, are located in quadrant II and III where $\Phi < 0$.
We therefore deduce that the closure of the hexapeptide along the blue pathway does not lead to a significant rotation of the Glycine torsion angles.
However, one orange cluster ($\chi \approx 0.65$) is located in quadrant IV ($\Phi \approx 0.95$), indicating that the red pathway involves a wide rotation of the $\Phi$ torsion angle of the Glycine: first about 120 degrees anti-clockwise, then again about 110 degrees clockwise.
We conclude from Fig.~\ref{fig:fig5}-(E) that this second pathway is more energy-intensive and, consequently, less probable.

\subsection*{Villin headpiece subdomain}
\begin{figure*}[bt!]
    \centering
    \includegraphics[width=1\textwidth]{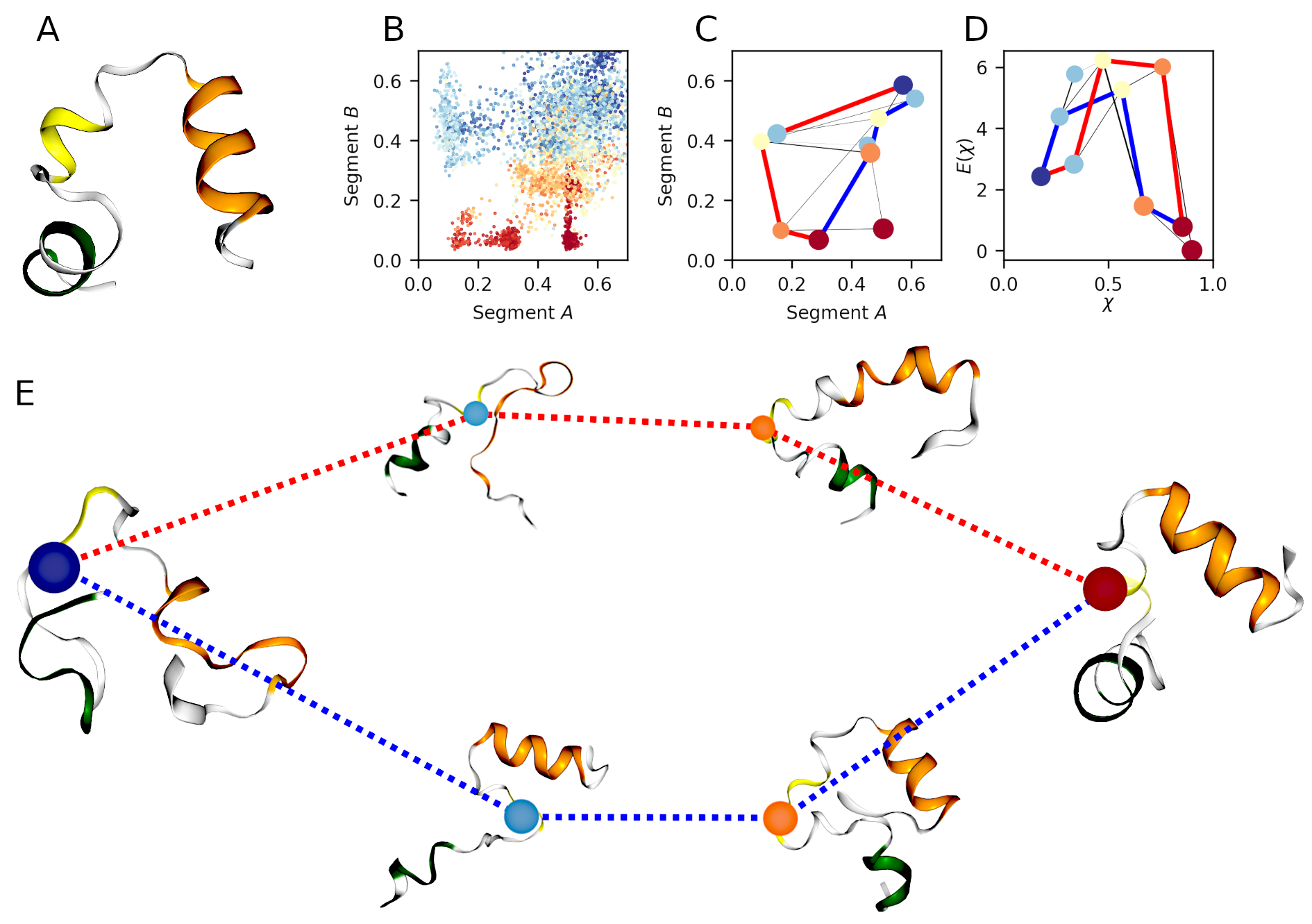}
    \caption{
    Results of villin headpiece subdomain.
    (A) The X-ray crystal structure of villin headpiece solved at pH 6.7, green, yellow and orange colors identify the helix $H_1$ (green), $H_2$ (yellow) and $H_3$ (orange) respectively; 
    (B) $\chi$-function projected onto the RMSD of segments $A$ and $B$.
    (C) MKM with representative structures  of the villin protein;
    (D) MKM projected  onto the RMSD of segments $A$ and $B$ of the villin protein;
    (E) Energy landscape of the villin protein with the most relevant clusters of the MKM.
    }
    \label{fig:fig6}
\end{figure*}
As a last example, to demonstrate the applicability of our approach to large systems, we studied the villin headpiece subdomain \cite{vardar2002villin,tang2006multistate,Lei2007,lindorff2011fast} which is one of the most studied protein for understanding protein folding.
Villin consists of 35 residues (582 atoms), and in its folded structure, it forms 3 $\alpha$-helices as shown in Fig.~\ref{fig:fig6}-(A): residues 4-8 form helix $H_1$ (green), residues 15-18 form helix $H_2$ (yellow), residues 23-32 form helix $H_3$ (orange).
In order to have a 2-dimensional representation of the molecular system, we used the RMSD of the segment $A$ (residues 3–21), which includes $H_1$ and $H_2$, and segment $B$ (residues 15–33), which includes $H_2$ and $H_3$, with respect to the corresponding segments of the X-ray crystal structure solved at pH 6.7 deposited in the RCSB protein data bank repository (PDB ID: 1YRF \cite{chiu2005high}) as done in Ref.~\cite{Lei2007}. 
Thus, a structure with low RMSD values corresponds to a folded structure, a structure with high RMSD values is an unfolded structure, and partially folded structures correspond to a situation where only the RMSD of one segment has low values. \\

\noindent \textbf{State space exploration and dynamics propagation.}
The folding timescale of the villin protein is $2.8\,\mathrm{\mu s}$, but a complete exploration of the space of states requires a conventional MD simulation of more than $2.8\,\mathrm{\mu s}$ \cite{lindorff2011fast}.
Alternatively, we carried out STMD simulations utilizing the dedicated OpenMM module for exploring the state space and selecting representative structures.
First, we prepared a complete extended structure with PyMol \cite{PyMOL}.
Then, we minimized the structure and equilibrated the system for 20 ps reaching a partially folded structure.
At this point, we carried out 6 independent replicas of $1\, \mathrm{\mu s}$ with temperatures ranging from 273 K to 500 K.
All the other parameters and options were as in the previous examples, the box contained 2713 water molecules.
From each replica, we extracted 1000 structures, for a total of 6000 structures, which constitute the set $X_0$ of initial states.
However, since temperature was a dynamic variable, we further equilibrated the structures for 100 ps in order to have a sample representing the Canonical Ensemble at $T=300\, \mathrm{K}$.
Afterward, we ran 10 short MD simulations of 1000 timesteps (0.2 ps) for each initial state.\\

\noindent \textbf{ISOKANN.}
The ISOKANN algorithm was applied as before, but we changed the input coordinates.
Indeed, since the system has 582 atoms, the number of pairwise distances is $\nicefrac{582\cdot (582- 1)}{2}=169071$.
Modern neural networks are able to handle this dimensionality, however, as a matter of efficiency and to show the versatility of the method, we preferred to reduce the dimensionality by using the internal coordinates (bonds, angles and torsion angles) of the backbone (140 atoms).
In this way, we reduced the number of dimensions to 1716.
We used a neural network with four layers (1716,  858,  429, 1), the initial learning rate of the SGD algorithm was 0.01 and the weight decay 0.005.
Convergence of the training occurred in 95 iterations with a training and validation loss in the order of $10^{-3}$ (SI Appendix, Fig.~S4).

In Fig.~\ref{fig:fig6}-(B), we report the $\chi$-function projected onto the two collective variables.
The bottom left corner (small values of RMSDs), contains folded structures with $\chi\approx 1$ (red).
As we move away from the corner, we observe transition structures with $\chi\approx 0.5$ (yellow) up to a large area containing unfolded structures with $\chi\approx 0$ (blue).\\

\noindent \textbf{MKM construction.}
We divided the $\chi$-function into 5 equal intervals and determined the optimal parameters $\varepsilon=0.9,  0.5,    0.3,    0.5,   0.5$ and $\theta=10,    60,     50,     150,   20$ for the CNN clustering algorithm and $r_n=0.6$ to determine the edges.
Thus we found 10 clusters, two clusters correspond to fully folded and unfolded structures, while the others contain partially folded structures.
The MKM is shown in Fig.~\ref{fig:fig6} in several representations: Fig.~\ref{fig:fig6}-(C) shows the projection onto the RMSDs of the segments $A$ and $B$;  Fig.~\ref{fig:fig6}-(D) shows the energy diagram as function of the $\chi$-values.  
(E) shows the edges between the most relevant representative structures of the protein (omitting the less relevant clusters of the MKM);\\

\noindent \textbf{Observations.}
We identify multiple folding pathways in the transition from unfolded to folded state, each with a distinct energetic profile. 
We have highlighted in blue the pathway that requires less energy, and in red the one that requires more, then the blue pathway is the most likely folding route.
This pathway corresponds to a scenario in which helix $H_3$ reaches its folded state more rapidly than the other two helices.
Conversely, the red pathway exhibits a process in which helix $H_1$ stabilizes prior to helices $H_2$ and $H_3$.
In both cases, the formation of helix $H_2$ is the slowest process and it starts at $\chi>0.5$.
%
%
%
\section*{Discussion}
\label{sec5}
The dynamics of a molecular system are highly complex and can be represented as a network of pathways connecting clusters of similar configurational states.
This happens for example in Markov State Models (MSMs) \cite{Prinz2011, Pande2018, Bowman2014, keller2018} where MD data are clustered and kinetics is described by transition probability matrices.
However, traditional clustering algorithms assume that data has certain shape or distribution, such as spherical in $k$-means or Gaussian in Gaussian Mixture Models.
This does not apply correctly to the high-dimensional state space of molecular systems which is characterized by complex geometries and non-trivial cluster shapes.
Another common problem when clustering MD simulations is the sparsity of transitional states, which could be mistakenly assigned to clusters representing metastable states instead of being interpreted as standalone transition clusters.
To alleviate this problem, we developed MoKiTo, an approach inspired by the Mapper algorithm, that does not make strict assumptions about cluster shapes and is more effective in capturing the inherent geometry and topology of the data. 
The idea behind MoKiTo is that molecular dynamics can be decomposed into dynamic processes, known also as Koopman modes, associated with different relaxation time scales.
Among these, the process associated with the largest non-zero eigenvalue represents long-term transitions, such as protein folding/unfolding or biomolecular binding/unbinding events, and indicates that the molecular system evolves over time along a preferred direction, characterized by the second eigenvector $\psi_1$ of the Koopman operator. 
Then, we use the $\chi$-function, linked to $\psi_1$ via eq.~\ref{eq:chi01}, to order the MD data according to their macroscopic features, facilitating subsequent clustering based on structural similarities.

The first example is particularly useful in showing the power of MoKiTo.
The potential energy function has four metastable states and by means of the $\chi$-function we identify three macroscopic regions, two metastable regions for corners $C_1$ and $C_4$ respectively, and one transition region that includes both corners $C_2$ and $C_3$.
Although it would be possible to cluster the initial states $X_0$ without using the $\chi$-function, this ordering function facilitates the clustering of states with similar macroscopic properties by quantifying how far the intermediate states are away from the starting state ($\chi$-value 0) or close to the end state ($\chi$-value 1).

The other examples show that MoKiTo also applies well to molecular systems of several orders of size.
Dichloroisobutene is a small molecule with two metastable states and two possible pathways, the clockwise and counterclockwise rotation of the torsion angle $\Psi$.
MoKiTo captures these properties and reveals that the red configurational states in Fig.~\ref{fig:fig4}-(D), i.e. with both chlorine atoms staggered with the methylene group, are the most stable states of the system. 
This is the expected result, since strain is minimized by distancing chlorines and the methyl group, which is more sterically demanding than the methylene group due to an additional hydrogen.

The third example shows that MoKiTo can be used to assess the quality of a reaction coordinate.
The dynamics of VGVAPG are characterized by the opening and closing of the salt-bridge between the positively charged N-terminus and the negatively charged C-terminus.
Intuitively, one might think that the distance between the extreme atoms of the peptide is a sufficient reaction coordinate.
Instead, via MoKiTo we revealed a more complex dynamic involving two possible rotations of the second residue of the hexapeptide.

Finally, the fourth example shows how to use MoKiTo to identify pathways in the folding/unfolding process of a protein.
In contrast to previous examples, where we used conventional MD simulations to sample the representative states of the state space, we used STMD simulations.
Indeed, the key requirement for MoKiTo is to have a broad representative set of states regardless of how these were sampled.
On the other hand, to estimate the $\chi$-function, the short trajectories should respect the true dynamics of the system, but it is not necessary that these trajectories reach a state of thermodynamic equilibrium.
We have identified two main pathways from the unfolded to folded state. 
In the first one (blue in Fig.~\ref{fig:fig6}) helix $H_3$ bends faster than the other two, while in the second one (red) helix $H_1$ bends faster. 
Of these, the first pathway is the most likely to occur as it crosses lower energy barriers. 
These results are consistent with previous findings in Refs.~\cite{Lei2007,Harada2012}. 
Also, mixed pathways appear from the MKM in Fig.~\ref{fig:fig6}, where helices form cooperatively at similar timescales, confirming the more recent findings in Ref.~\cite{Wang2019}.

Protein misfolding is a major contributor to various diseases and is extensively researched using MD simulations. 
The existence of numerous potential pathways makes it challenging for standard MD techniques and analytical tools to effectively capture this multiplicity. 
This example demonstrates that MoKiTo directly addresses this issue, allowing diversity in these pathways to be studied.

\section*{Conclusion}
\label{sec6}
In summary, MoKiTo is an analysis tool that captures molecular kinetics using graph representations to elucidate conformational changes and provides
a complete one-dimensional representation of the system that allows the energy levels of the clusters and the energy cost of the paths to be quantified.
Unlike traditional methods, MoKiTo does not rely on long MD simulations to track the system's time evolution. 
Instead, it requires only a representative sampling of the state space and an ordering function, $\chi$, that defines the ``forward'' and ``backward''  directions of progression.
This approach bypasses assumptions about cluster shapes or state distributions, allowing for a more accurate representation of the system's topology.
The $\chi$-function was determined using ISOKANN, which is advantageous because it can be applied to short trajectories that do not reach thermal equilibrium and does not require prior definition of metastable macrostates as discrete state space subsets, eliminating the need for detailed prior knowledge of the system.
Additionally, while MKMs reveal the structure of multiple pathways, a single reaction coordinate, $\chi$, suffices to represent the entire process. 
This idea is illustrated in Fig.~\ref{fig:fig1}-(B) where there are five distinct fingers, but one ordering parameter is enough to capture the complete topology of the hand.

%% file: 0_main_preprint.bbl
\begin{thebibliography}{48}%
\makeatletter
\providecommand \@ifxundefined [1]{%
 \@ifx{#1\undefined}
}%
\providecommand \@ifnum [1]{%
 \ifnum #1\expandafter \@firstoftwo
 \else \expandafter \@secondoftwo
 \fi
}%
\providecommand \@ifx [1]{%
 \ifx #1\expandafter \@firstoftwo
 \else \expandafter \@secondoftwo
 \fi
}%
\providecommand \natexlab [1]{#1}%
\providecommand \enquote  [1]{``#1''}%
\providecommand \bibnamefont  [1]{#1}%
\providecommand \bibfnamefont [1]{#1}%
\providecommand \citenamefont [1]{#1}%
\providecommand \href@noop [0]{\@secondoftwo}%
\providecommand \href [0]{\begingroup \@sanitize@url \@href}%
\providecommand \@href[1]{\@@startlink{#1}\@@href}%
\providecommand \@@href[1]{\endgroup#1\@@endlink}%
\providecommand \@sanitize@url [0]{\catcode `\\12\catcode `\$12\catcode `\&12\catcode `\#12\catcode `\^12\catcode `\_12\catcode `\%12\relax}%
\providecommand \@@startlink[1]{}%
\providecommand \@@endlink[0]{}%
\providecommand \url  [0]{\begingroup\@sanitize@url \@url }%
\providecommand \@url [1]{\endgroup\@href {#1}{\urlprefix }}%
\providecommand \urlprefix  [0]{URL }%
\providecommand \Eprint [0]{\href }%
\providecommand \doibase [0]{http://dx.doi.org/}%
\providecommand \selectlanguage [0]{\@gobble}%
\providecommand \bibinfo  [0]{\@secondoftwo}%
\providecommand \bibfield  [0]{\@secondoftwo}%
\providecommand \translation [1]{[#1]}%
\providecommand \BibitemOpen [0]{}%
\providecommand \bibitemStop [0]{}%
\providecommand \bibitemNoStop [0]{.\EOS\space}%
\providecommand \EOS [0]{\spacefactor3000\relax}%
\providecommand \BibitemShut  [1]{\csname bibitem#1\endcsname}%
\let\auto@bib@innerbib\@empty
\bibitem [{\citenamefont {Karplus}\ and\ \citenamefont {McCammon}(2002)}]{karplus2002molecular}%
  \BibitemOpen
  \bibfield  {author} {\bibinfo {author} {\bibfnamefont {M.}~\bibnamefont {Karplus}}\ and\ \bibinfo {author} {\bibfnamefont {J.~A.}\ \bibnamefont {McCammon}},\ }\bibfield  {title} {\enquote {\bibinfo {title} {Molecular dynamics simulations of biomolecules},}\ }\href@noop {} {\bibfield  {journal} {\bibinfo  {journal} {Nature Structural Biology}\ }\textbf {\bibinfo {volume} {9}},\ \bibinfo {pages} {646--652} (\bibinfo {year} {2002})}\BibitemShut {NoStop}%
\bibitem [{\citenamefont {Frenkel}\ and\ \citenamefont {Smit}(2001)}]{frenkel2001understanding}%
  \BibitemOpen
  \bibfield  {author} {\bibinfo {author} {\bibfnamefont {D.}~\bibnamefont {Frenkel}}\ and\ \bibinfo {author} {\bibfnamefont {B.}~\bibnamefont {Smit}},\ }\href@noop {} {\emph {\bibinfo {title} {Understanding Molecular Simulation: From Algorithms to Applications}}}\ (\bibinfo  {publisher} {Academic Press},\ \bibinfo {year} {2001})\BibitemShut {NoStop}%
\bibitem [{\citenamefont {Allen}\ and\ \citenamefont {Tildesley}(2017)}]{allen2017computer}%
  \BibitemOpen
  \bibfield  {author} {\bibinfo {author} {\bibfnamefont {M.~P.}\ \bibnamefont {Allen}}\ and\ \bibinfo {author} {\bibfnamefont {D.~J.}\ \bibnamefont {Tildesley}},\ }\href@noop {} {\emph {\bibinfo {title} {Computer Simulation of Liquids}}}\ (\bibinfo  {publisher} {Oxford University Press},\ \bibinfo {year} {2017})\BibitemShut {NoStop}%
\bibitem [{\citenamefont {Laio}\ and\ \citenamefont {Parrinello}(2002{\natexlab{a}})}]{laio2002escaping}%
  \BibitemOpen
  \bibfield  {author} {\bibinfo {author} {\bibfnamefont {A.}~\bibnamefont {Laio}}\ and\ \bibinfo {author} {\bibfnamefont {M.}~\bibnamefont {Parrinello}},\ }\bibfield  {title} {\enquote {\bibinfo {title} {Escaping free-energy minima},}\ }\href@noop {} {\bibfield  {journal} {\bibinfo  {journal} {Proceedings of the National Academy of Sciences}\ }\textbf {\bibinfo {volume} {99}},\ \bibinfo {pages} {12562--12566} (\bibinfo {year} {2002}{\natexlab{a}})}\BibitemShut {NoStop}%
\bibitem [{\citenamefont {Dellago}, \citenamefont {Bolhuis},\ and\ \citenamefont {Chandler}(1998)}]{dellago1998efficient}%
  \BibitemOpen
  \bibfield  {author} {\bibinfo {author} {\bibfnamefont {C.}~\bibnamefont {Dellago}}, \bibinfo {author} {\bibfnamefont {P.~G.}\ \bibnamefont {Bolhuis}}, \ and\ \bibinfo {author} {\bibfnamefont {D.}~\bibnamefont {Chandler}},\ }\bibfield  {title} {\enquote {\bibinfo {title} {Efficient transition path sampling: Application to lennard-jones cluster rearrangements},}\ }\href@noop {} {\bibfield  {journal} {\bibinfo  {journal} {The Journal of Chemical Physics}\ }\textbf {\bibinfo {volume} {108}},\ \bibinfo {pages} {9236--9245} (\bibinfo {year} {1998})}\BibitemShut {NoStop}%
\bibitem [{\citenamefont {Deuflhard}\ and\ \citenamefont {Weber}(2004)}]{Deuflhard2004}%
  \BibitemOpen
  \bibfield  {author} {\bibinfo {author} {\bibfnamefont {P.}~\bibnamefont {Deuflhard}}\ and\ \bibinfo {author} {\bibfnamefont {M.}~\bibnamefont {Weber}},\ }\bibfield  {title} {\enquote {\bibinfo {title} {{Robust Perron cluster analysis in conformation dynamics}},}\ }\href@noop {} {\bibfield  {journal} {\bibinfo  {journal} {Linear Algebra Appl.}\ }\textbf {\bibinfo {volume} {398}},\ \bibinfo {pages} {161--184} (\bibinfo {year} {2004})}\BibitemShut {NoStop}%
\bibitem [{\citenamefont {Weber}(2006)}]{Weber2006thesis}%
  \BibitemOpen
  \bibfield  {author} {\bibinfo {author} {\bibfnamefont {M.}~\bibnamefont {Weber}},\ }\emph {\bibinfo {title} {Meshless Methods in Conformation Dynamics}},\ \href@noop {} {Ph.D. thesis},\ \bibinfo  {school} {FU Berlin} (\bibinfo {year} {2006})\BibitemShut {NoStop}%
\bibitem [{\citenamefont {Rabben}, \citenamefont {Ray},\ and\ \citenamefont {Weber}(2020)}]{Rabben2020}%
  \BibitemOpen
  \bibfield  {author} {\bibinfo {author} {\bibfnamefont {R.~J.}\ \bibnamefont {Rabben}}, \bibinfo {author} {\bibfnamefont {S.}~\bibnamefont {Ray}}, \ and\ \bibinfo {author} {\bibfnamefont {M.}~\bibnamefont {Weber}},\ }\bibfield  {title} {\enquote {\bibinfo {title} {{ISOKANN: Invariant subspaces of Koopman operators learned by a neural network}},}\ }\href {\doibase 10.1063/5.0015132} {\bibfield  {journal} {\bibinfo  {journal} {J. Chem. Phys.}\ }\textbf {\bibinfo {volume} {153}},\ \bibinfo {pages} {114109} (\bibinfo {year} {2020})}\BibitemShut {NoStop}%
\bibitem [{\citenamefont {Sikorski}, \citenamefont {Ribera~Borrell},\ and\ \citenamefont {Weber}(2024)}]{Sikorski2024}%
  \BibitemOpen
  \bibfield  {author} {\bibinfo {author} {\bibfnamefont {A.}~\bibnamefont {Sikorski}}, \bibinfo {author} {\bibfnamefont {E.}~\bibnamefont {Ribera~Borrell}}, \ and\ \bibinfo {author} {\bibfnamefont {M.}~\bibnamefont {Weber}},\ }\bibfield  {title} {\enquote {\bibinfo {title} {Learning koopman eigenfunctions of stochastic diffusions with optimal importance sampling and isokann},}\ }\href {\doibase 10.1063/5.0140764} {\bibfield  {journal} {\bibinfo  {journal} {Journal of Mathematical Physics}\ }\textbf {\bibinfo {volume} {65}},\ \bibinfo {pages} {013502} (\bibinfo {year} {2024})}\BibitemShut {NoStop}%
\bibitem [{\citenamefont {Donati}, \citenamefont {Schütte},\ and\ \citenamefont {Weber}(2024)}]{Donati2024}%
  \BibitemOpen
  \bibfield  {author} {\bibinfo {author} {\bibfnamefont {L.}~\bibnamefont {Donati}}, \bibinfo {author} {\bibfnamefont {C.}~\bibnamefont {Schütte}}, \ and\ \bibinfo {author} {\bibfnamefont {M.}~\bibnamefont {Weber}},\ }\bibfield  {title} {\enquote {\bibinfo {title} {The kramers turnover in terms of a macro-state projection on phase space},}\ }\href {\doibase 10.1080/00268976.2024.2356748} {\bibfield  {journal} {\bibinfo  {journal} {Mol. Phys.}\ }\textbf {\bibinfo {volume} {0}},\ \bibinfo {pages} {e2356748} (\bibinfo {year} {2024})}\BibitemShut {NoStop}%
\bibitem [{\citenamefont {Weber}\ and\ \citenamefont {Ernst}(2017)}]{weber2017fuzzy}%
  \BibitemOpen
  \bibfield  {author} {\bibinfo {author} {\bibfnamefont {M.}~\bibnamefont {Weber}}\ and\ \bibinfo {author} {\bibfnamefont {N.}~\bibnamefont {Ernst}},\ }\bibfield  {title} {\enquote {\bibinfo {title} {A fuzzy-set theoretical framework for computing exit rates of rare events in potential-driven diffusion processes},}\ }\href@noop {} {\bibfield  {journal} {\bibinfo  {journal} {arXiv preprint arXiv:1708.00679}\ } (\bibinfo {year} {2017})}\BibitemShut {NoStop}%
\bibitem [{\citenamefont {Singh}, \citenamefont {Memoli},\ and\ \citenamefont {Carlsson}(2007)}]{Gurjeet2007}%
  \BibitemOpen
  \bibfield  {author} {\bibinfo {author} {\bibfnamefont {G.}~\bibnamefont {Singh}}, \bibinfo {author} {\bibfnamefont {F.}~\bibnamefont {Memoli}}, \ and\ \bibinfo {author} {\bibfnamefont {G.}~\bibnamefont {Carlsson}},\ }\bibfield  {title} {\enquote {\bibinfo {title} {{Topological Methods for the Analysis of High Dimensional Data Sets and 3D Object Recognition }},}\ }in\ \href {\doibase /10.2312/SPBG/SPBG07/091-100} {\emph {\bibinfo {booktitle} {Eurographics Symposium on Point-Based Graphics}}},\ \bibinfo {editor} {edited by\ \bibinfo {editor} {\bibfnamefont {M.}~\bibnamefont {Botsch}}, \bibinfo {editor} {\bibfnamefont {R.}~\bibnamefont {Pajarola}}, \bibinfo {editor} {\bibfnamefont {B.}~\bibnamefont {Chen}}, \ and\ \bibinfo {editor} {\bibfnamefont {M.}~\bibnamefont {Zwicker}}}\ (\bibinfo  {publisher} {The Eurographics Association},\ \bibinfo {year} {2007})\BibitemShut {NoStop}%
\bibitem [{\citenamefont {Lum}\ \emph {et~al.}(2013)\citenamefont {Lum}, \citenamefont {Singh}, \citenamefont {Lehman}, \citenamefont {Ishkanov}, \citenamefont {Vejdemo-Johansson}, \citenamefont {Alagappan}, \citenamefont {Carlsson},\ and\ \citenamefont {Carlsson}}]{Lum2013}%
  \BibitemOpen
  \bibfield  {author} {\bibinfo {author} {\bibfnamefont {P.}~\bibnamefont {Lum}}, \bibinfo {author} {\bibfnamefont {G.}~\bibnamefont {Singh}}, \bibinfo {author} {\bibfnamefont {A.}~\bibnamefont {Lehman}}, \bibinfo {author} {\bibfnamefont {T.}~\bibnamefont {Ishkanov}}, \bibinfo {author} {\bibfnamefont {M.}~\bibnamefont {Vejdemo-Johansson}}, \bibinfo {author} {\bibfnamefont {M.}~\bibnamefont {Alagappan}}, \bibinfo {author} {\bibfnamefont {J.}~\bibnamefont {Carlsson}}, \ and\ \bibinfo {author} {\bibfnamefont {G.}~\bibnamefont {Carlsson}},\ }\bibfield  {title} {\enquote {\bibinfo {title} {{Extracting insights from the shape of complex data using topology}},}\ }\href@noop {} {\bibfield  {journal} {\bibinfo  {journal} {Sci. Rep.}\ }\textbf {\bibinfo {volume} {3}} (\bibinfo {year} {2013})}\BibitemShut {NoStop}%
\bibitem [{\citenamefont {Vardar}\ \emph {et~al.}(2002)\citenamefont {Vardar}, \citenamefont {Chishti}, \citenamefont {Frank}, \citenamefont {Luna}, \citenamefont {Noegel}, \citenamefont {Oh}, \citenamefont {Schleicher},\ and\ \citenamefont {McKnight}}]{vardar2002villin}%
  \BibitemOpen
  \bibfield  {author} {\bibinfo {author} {\bibfnamefont {D.}~\bibnamefont {Vardar}}, \bibinfo {author} {\bibfnamefont {A.}~\bibnamefont {Chishti}}, \bibinfo {author} {\bibfnamefont {B.}~\bibnamefont {Frank}}, \bibinfo {author} {\bibfnamefont {E.~J.}\ \bibnamefont {Luna}}, \bibinfo {author} {\bibfnamefont {A.}~\bibnamefont {Noegel}}, \bibinfo {author} {\bibfnamefont {S.~W.}\ \bibnamefont {Oh}}, \bibinfo {author} {\bibfnamefont {M.}~\bibnamefont {Schleicher}}, \ and\ \bibinfo {author} {\bibfnamefont {C.}~\bibnamefont {McKnight}},\ }\bibfield  {title} {\enquote {\bibinfo {title} {Villin-type headpiece domains show a wide range of f-actin-binding affinities},}\ }\href@noop {} {\bibfield  {journal} {\bibinfo  {journal} {Cell motility and the cytoskeleton}\ }\textbf {\bibinfo {volume} {52}},\ \bibinfo {pages} {9--21} (\bibinfo {year} {2002})}\BibitemShut {NoStop}%
\bibitem [{\citenamefont {Tang}\ \emph {et~al.}(2006)\citenamefont {Tang}, \citenamefont {Grey}, \citenamefont {McKnight}, \citenamefont {Palmer~III},\ and\ \citenamefont {Raleigh}}]{tang2006multistate}%
  \BibitemOpen
  \bibfield  {author} {\bibinfo {author} {\bibfnamefont {Y.}~\bibnamefont {Tang}}, \bibinfo {author} {\bibfnamefont {M.~J.}\ \bibnamefont {Grey}}, \bibinfo {author} {\bibfnamefont {J.}~\bibnamefont {McKnight}}, \bibinfo {author} {\bibfnamefont {A.~G.}\ \bibnamefont {Palmer~III}}, \ and\ \bibinfo {author} {\bibfnamefont {D.~P.}\ \bibnamefont {Raleigh}},\ }\bibfield  {title} {\enquote {\bibinfo {title} {Multistate folding of the villin headpiece domain},}\ }\href@noop {} {\bibfield  {journal} {\bibinfo  {journal} {Journal of molecular biology}\ }\textbf {\bibinfo {volume} {355}},\ \bibinfo {pages} {1066--1077} (\bibinfo {year} {2006})}\BibitemShut {NoStop}%
\bibitem [{\citenamefont {Lei}\ \emph {et~al.}(2007)\citenamefont {Lei}, \citenamefont {Wu}, \citenamefont {Liu},\ and\ \citenamefont {Duan}}]{Lei2007}%
  \BibitemOpen
  \bibfield  {author} {\bibinfo {author} {\bibfnamefont {H.}~\bibnamefont {Lei}}, \bibinfo {author} {\bibfnamefont {C.}~\bibnamefont {Wu}}, \bibinfo {author} {\bibfnamefont {H.}~\bibnamefont {Liu}}, \ and\ \bibinfo {author} {\bibfnamefont {Y.}~\bibnamefont {Duan}},\ }\bibfield  {title} {\enquote {\bibinfo {title} {Folding free-energy landscape of villin headpiece subdomain from molecular dynamics simulations},}\ }\href {\doibase 10.1073/pnas.0608432104} {\bibfield  {journal} {\bibinfo  {journal} {Proceedings of the National Academy of Sciences of the United States of America}\ }\textbf {\bibinfo {volume} {104}},\ \bibinfo {pages} {4925--30} (\bibinfo {year} {2007})}\BibitemShut {NoStop}%
\bibitem [{\citenamefont {Lindorff-Larsen}\ \emph {et~al.}(2011)\citenamefont {Lindorff-Larsen}, \citenamefont {Piana}, \citenamefont {Dror},\ and\ \citenamefont {Shaw}}]{lindorff2011fast}%
  \BibitemOpen
  \bibfield  {author} {\bibinfo {author} {\bibfnamefont {K.}~\bibnamefont {Lindorff-Larsen}}, \bibinfo {author} {\bibfnamefont {S.}~\bibnamefont {Piana}}, \bibinfo {author} {\bibfnamefont {R.~O.}\ \bibnamefont {Dror}}, \ and\ \bibinfo {author} {\bibfnamefont {D.~E.}\ \bibnamefont {Shaw}},\ }\bibfield  {title} {\enquote {\bibinfo {title} {How fast-folding proteins fold},}\ }\href@noop {} {\bibfield  {journal} {\bibinfo  {journal} {Science}\ }\textbf {\bibinfo {volume} {334}},\ \bibinfo {pages} {517--520} (\bibinfo {year} {2011})}\BibitemShut {NoStop}%
\bibitem [{\citenamefont {Marinari}\ and\ \citenamefont {Parisi}(1992)}]{Marinari_1992}%
  \BibitemOpen
  \bibfield  {author} {\bibinfo {author} {\bibfnamefont {E.}~\bibnamefont {Marinari}}\ and\ \bibinfo {author} {\bibfnamefont {G.}~\bibnamefont {Parisi}},\ }\bibfield  {title} {\enquote {\bibinfo {title} {Simulated tempering: A new monte carlo scheme},}\ }\href {\doibase 10.1209/0295-5075/19/6/002} {\bibfield  {journal} {\bibinfo  {journal} {Europhysics Letters}\ }\textbf {\bibinfo {volume} {19}},\ \bibinfo {pages} {451} (\bibinfo {year} {1992})}\BibitemShut {NoStop}%
\bibitem [{\citenamefont {Sugita}\ and\ \citenamefont {Okamoto}(2000)}]{Sugita2000}%
  \BibitemOpen
  \bibfield  {author} {\bibinfo {author} {\bibfnamefont {Y.}~\bibnamefont {Sugita}}\ and\ \bibinfo {author} {\bibfnamefont {Y.}~\bibnamefont {Okamoto}},\ }\bibfield  {title} {\enquote {\bibinfo {title} {{Replica-exchange multicanonical algorithm and multicanonical replica-exchange method for simulating systems with rough energy landscape}},}\ }\href@noop {} {\bibfield  {journal} {\bibinfo  {journal} {Chem. Phys. Lett.}\ }\textbf {\bibinfo {volume} {329}},\ \bibinfo {pages} {261--270} (\bibinfo {year} {2000})}\BibitemShut {NoStop}%
\bibitem [{\citenamefont {Souaille}\ and\ \citenamefont {Roux}(2001)}]{Souaille2001}%
  \BibitemOpen
  \bibfield  {author} {\bibinfo {author} {\bibfnamefont {M.}~\bibnamefont {Souaille}}\ and\ \bibinfo {author} {\bibfnamefont {B.}~\bibnamefont {Roux}},\ }\bibfield  {title} {\enquote {\bibinfo {title} {{Extension to the weighted histogram analysis method: combining umbrella sampling with free energy calculations}},}\ }\href {\doibase 10.1002/jcc.540130812} {\bibfield  {journal} {\bibinfo  {journal} {Comp. Phys- Comm.}\ }\textbf {\bibinfo {volume} {135}},\ \bibinfo {pages} {40–57} (\bibinfo {year} {2001})}\BibitemShut {NoStop}%
\bibitem [{\citenamefont {Huber}, \citenamefont {Torda},\ and\ \citenamefont {van Gunsteren}(1994)}]{Huber1994}%
  \BibitemOpen
  \bibfield  {author} {\bibinfo {author} {\bibfnamefont {T.}~\bibnamefont {Huber}}, \bibinfo {author} {\bibfnamefont {A.}~\bibnamefont {Torda}}, \ and\ \bibinfo {author} {\bibfnamefont {W.}~\bibnamefont {van Gunsteren}},\ }\bibfield  {title} {\enquote {\bibinfo {title} {{Local elevation: A method for improving the searching properties of molecular dynamics simulation}},}\ }\href@noop {} {\bibfield  {journal} {\bibinfo  {journal} {J. Comput. Aided Mol. Des.}\ }\textbf {\bibinfo {volume} {8}} (\bibinfo {year} {1994})}\BibitemShut {NoStop}%
\bibitem [{\citenamefont {Laio}\ and\ \citenamefont {Parrinello}(2002{\natexlab{b}})}]{Laio2002}%
  \BibitemOpen
  \bibfield  {author} {\bibinfo {author} {\bibfnamefont {A.}~\bibnamefont {Laio}}\ and\ \bibinfo {author} {\bibfnamefont {M.}~\bibnamefont {Parrinello}},\ }\bibfield  {title} {\enquote {\bibinfo {title} {Escaping free-energy minima.}}\ }\href@noop {} {\bibfield  {journal} {\bibinfo  {journal} {Proc. Natl. Acad. Sci. U.S.A.}\ }\textbf {\bibinfo {volume} {99}},\ \bibinfo {pages} {12562--6} (\bibinfo {year} {2002}{\natexlab{b}})}\BibitemShut {NoStop}%
\bibitem [{\citenamefont {Barducci}, \citenamefont {Bussi},\ and\ \citenamefont {Parrinello}(2008)}]{Barducci2008}%
  \BibitemOpen
  \bibfield  {author} {\bibinfo {author} {\bibfnamefont {A.}~\bibnamefont {Barducci}}, \bibinfo {author} {\bibfnamefont {G.}~\bibnamefont {Bussi}}, \ and\ \bibinfo {author} {\bibfnamefont {M.}~\bibnamefont {Parrinello}},\ }\bibfield  {title} {\enquote {\bibinfo {title} {Well-tempered metadynamics: A smoothly converging and tunable free-energy method.}}\ }\href@noop {} {\bibfield  {journal} {\bibinfo  {journal} {Phys. Rev. Lett.}\ }\textbf {\bibinfo {volume} {100}},\ \bibinfo {pages} {020603} (\bibinfo {year} {2008})}\BibitemShut {NoStop}%
\bibitem [{\citenamefont {von Mises}\ and\ \citenamefont {Pollaczek-Geiringer}(1929)}]{Mises1929}%
  \BibitemOpen
  \bibfield  {author} {\bibinfo {author} {\bibfnamefont {R.}~\bibnamefont {von Mises}}\ and\ \bibinfo {author} {\bibfnamefont {H.}~\bibnamefont {Pollaczek-Geiringer}},\ }\bibfield  {title} {\enquote {\bibinfo {title} {Praktische verfahren der gleichungsaufl{\"o}sung .}}\ }\href {https://api.semanticscholar.org/CorpusID:120916484} {\bibfield  {journal} {\bibinfo  {journal} {Zamm-zeitschrift Fur Angewandte Mathematik Und Mechanik}\ }\textbf {\bibinfo {volume} {9}},\ \bibinfo {pages} {152--164} (\bibinfo {year} {1929})}\BibitemShut {NoStop}%
\bibitem [{\citenamefont {Weber}(2018)}]{Weber2018}%
  \BibitemOpen
  \bibfield  {author} {\bibinfo {author} {\bibfnamefont {M.}~\bibnamefont {Weber}},\ }\bibfield  {title} {\enquote {\bibinfo {title} {{Implications of PCCA+ in Molecular Simulation}},}\ }\href@noop {} {\bibfield  {journal} {\bibinfo  {journal} {Computation}\ }\textbf {\bibinfo {volume} {6}} (\bibinfo {year} {2018})}\BibitemShut {NoStop}%
\bibitem [{\citenamefont {Keller}, \citenamefont {Daura},\ and\ \citenamefont {van Gunsteren}(2010)}]{Keller2010}%
  \BibitemOpen
  \bibfield  {author} {\bibinfo {author} {\bibfnamefont {B.}~\bibnamefont {Keller}}, \bibinfo {author} {\bibfnamefont {X.}~\bibnamefont {Daura}}, \ and\ \bibinfo {author} {\bibfnamefont {W.~F.}\ \bibnamefont {van Gunsteren}},\ }\bibfield  {title} {\enquote {\bibinfo {title} {{Comparing geometric and kinetic cluster algorithms for molecular simulation data}},}\ }\href {\doibase 10.1063/1.3301140} {\bibfield  {journal} {\bibinfo  {journal} {J. Chem. Phys.}\ }\textbf {\bibinfo {volume} {132}},\ \bibinfo {pages} {074110} (\bibinfo {year} {2010})}\BibitemShut {NoStop}%
\bibitem [{\citenamefont {Lemke}\ and\ \citenamefont {Keller}(2016)}]{Lemke2016}%
  \BibitemOpen
  \bibfield  {author} {\bibinfo {author} {\bibfnamefont {O.}~\bibnamefont {Lemke}}\ and\ \bibinfo {author} {\bibfnamefont {B.~G.}\ \bibnamefont {Keller}},\ }\bibfield  {title} {\enquote {\bibinfo {title} {{Density-based cluster algorithms for the identification of core sets}},}\ }\href {\doibase 10.1063/1.4965440} {\bibfield  {journal} {\bibinfo  {journal} {The Journal of Chemical Physics}\ }\textbf {\bibinfo {volume} {145}},\ \bibinfo {pages} {164104} (\bibinfo {year} {2016})}\BibitemShut {NoStop}%
\bibitem [{\citenamefont {Lemke}\ and\ \citenamefont {Keller}(2018)}]{Lemke2019}%
  \BibitemOpen
  \bibfield  {author} {\bibinfo {author} {\bibfnamefont {O.}~\bibnamefont {Lemke}}\ and\ \bibinfo {author} {\bibfnamefont {B.~G.}\ \bibnamefont {Keller}},\ }\bibfield  {title} {\enquote {\bibinfo {title} {Common nearest neighbor clustering—a benchmark},}\ }\href {\doibase 10.3390/a11020019} {\bibfield  {journal} {\bibinfo  {journal} {Algorithms}\ }\textbf {\bibinfo {volume} {11}} (\bibinfo {year} {2018}),\ 10.3390/a11020019}\BibitemShut {NoStop}%
\bibitem [{\citenamefont {Leimkuhler}\ and\ \citenamefont {Matthews}(2015)}]{Leimkuhler2015}%
  \BibitemOpen
  \bibfield  {author} {\bibinfo {author} {\bibfnamefont {B.}~\bibnamefont {Leimkuhler}}\ and\ \bibinfo {author} {\bibfnamefont {C.}~\bibnamefont {Matthews}},\ }\href@noop {} {\emph {\bibinfo {title} {Molecular Dynamics: With Deterministic and Stochastic Numerical Methods.}}}\ (\bibinfo  {publisher} {Springer},\ \bibinfo {address} {Interdisciplinary Applied Mathematics; Vol. 39},\ \bibinfo {year} {2015})\BibitemShut {NoStop}%
\bibitem [{\citenamefont {Paszke}\ \emph {et~al.}(2019)\citenamefont {Paszke}, \citenamefont {Gross}, \citenamefont {Massa}, \citenamefont {Lerer}, \citenamefont {Bradbury}, \citenamefont {Chanan}, \citenamefont {Killeen}, \citenamefont {Lin}, \citenamefont {Gimelshein}, \citenamefont {Antiga}, \citenamefont {Desmaison}, \citenamefont {Kopf}, \citenamefont {Yang}, \citenamefont {DeVito}, \citenamefont {Raison}, \citenamefont {Tejani}, \citenamefont {Chilamkurthy}, \citenamefont {Steiner}, \citenamefont {Fang}, \citenamefont {Bai},\ and\ \citenamefont {Chintala}}]{NEURIPS2019_9015}%
  \BibitemOpen
  \bibfield  {author} {\bibinfo {author} {\bibfnamefont {A.}~\bibnamefont {Paszke}}, \bibinfo {author} {\bibfnamefont {S.}~\bibnamefont {Gross}}, \bibinfo {author} {\bibfnamefont {F.}~\bibnamefont {Massa}}, \bibinfo {author} {\bibfnamefont {A.}~\bibnamefont {Lerer}}, \bibinfo {author} {\bibfnamefont {J.}~\bibnamefont {Bradbury}}, \bibinfo {author} {\bibfnamefont {G.}~\bibnamefont {Chanan}}, \bibinfo {author} {\bibfnamefont {T.}~\bibnamefont {Killeen}}, \bibinfo {author} {\bibfnamefont {Z.}~\bibnamefont {Lin}}, \bibinfo {author} {\bibfnamefont {N.}~\bibnamefont {Gimelshein}}, \bibinfo {author} {\bibfnamefont {L.}~\bibnamefont {Antiga}}, \bibinfo {author} {\bibfnamefont {A.}~\bibnamefont {Desmaison}}, \bibinfo {author} {\bibfnamefont {A.}~\bibnamefont {Kopf}}, \bibinfo {author} {\bibfnamefont {E.}~\bibnamefont {Yang}}, \bibinfo {author} {\bibfnamefont {Z.}~\bibnamefont {DeVito}}, \bibinfo {author} {\bibfnamefont {M.}~\bibnamefont {Raison}}, \bibinfo {author} {\bibfnamefont {A.}~\bibnamefont {Tejani}}, \bibinfo
  {author} {\bibfnamefont {S.}~\bibnamefont {Chilamkurthy}}, \bibinfo {author} {\bibfnamefont {B.}~\bibnamefont {Steiner}}, \bibinfo {author} {\bibfnamefont {L.}~\bibnamefont {Fang}}, \bibinfo {author} {\bibfnamefont {J.}~\bibnamefont {Bai}}, \ and\ \bibinfo {author} {\bibfnamefont {S.}~\bibnamefont {Chintala}},\ }\bibfield  {title} {\enquote {\bibinfo {title} {Pytorch: An imperative style, high-performance deep learning library},}\ }in\ \href {http://papers.neurips.cc/paper/9015-pytorch-an-imperative-style-high-performance-deep-learning-library.pdf} {\emph {\bibinfo {booktitle} {Advances in Neural Information Processing Systems 32}}}\ (\bibinfo  {publisher} {Curran Associates, Inc.},\ \bibinfo {year} {2019})\ pp.\ \bibinfo {pages} {8024--8035}\BibitemShut {NoStop}%
\bibitem [{\citenamefont {Robbins}(1951)}]{Robbins1951ASA}%
  \BibitemOpen
  \bibfield  {author} {\bibinfo {author} {\bibfnamefont {H.~E.}\ \bibnamefont {Robbins}},\ }\bibfield  {title} {\enquote {\bibinfo {title} {A stochastic approximation method},}\ }\href {https://api.semanticscholar.org/CorpusID:16945044} {\bibfield  {journal} {\bibinfo  {journal} {Annals of Mathematical Statistics}\ }\textbf {\bibinfo {volume} {22}},\ \bibinfo {pages} {400--407} (\bibinfo {year} {1951})}\BibitemShut {NoStop}%
\bibitem [{\citenamefont {Eastman}\ \emph {et~al.}(2017)\citenamefont {Eastman}, \citenamefont {Swails}, \citenamefont {Chodera}, \citenamefont {McGibbon}, \citenamefont {Zhao}, \citenamefont {Beauchamp}, \citenamefont {Wang}, \citenamefont {Simmonett}, \citenamefont {Harrigan}, \citenamefont {Stern}, \citenamefont {Wiewiora}, \citenamefont {Brooks},\ and\ \citenamefont {Pande}}]{OpenMM2017}%
  \BibitemOpen
  \bibfield  {author} {\bibinfo {author} {\bibfnamefont {P.}~\bibnamefont {Eastman}}, \bibinfo {author} {\bibfnamefont {J.}~\bibnamefont {Swails}}, \bibinfo {author} {\bibfnamefont {J.~D.}\ \bibnamefont {Chodera}}, \bibinfo {author} {\bibfnamefont {R.~T.}\ \bibnamefont {McGibbon}}, \bibinfo {author} {\bibfnamefont {Y.}~\bibnamefont {Zhao}}, \bibinfo {author} {\bibfnamefont {K.~A.}\ \bibnamefont {Beauchamp}}, \bibinfo {author} {\bibfnamefont {L.-P.}\ \bibnamefont {Wang}}, \bibinfo {author} {\bibfnamefont {A.~C.}\ \bibnamefont {Simmonett}}, \bibinfo {author} {\bibfnamefont {M.~P.}\ \bibnamefont {Harrigan}}, \bibinfo {author} {\bibfnamefont {C.~D.}\ \bibnamefont {Stern}}, \bibinfo {author} {\bibfnamefont {R.~P.}\ \bibnamefont {Wiewiora}}, \bibinfo {author} {\bibfnamefont {B.~R.}\ \bibnamefont {Brooks}}, \ and\ \bibinfo {author} {\bibfnamefont {V.~S.}\ \bibnamefont {Pande}},\ }\bibfield  {title} {\enquote {\bibinfo {title} {Openmm 7: Rapid development of high performance algorithms for molecular dynamics},}\
  }\href {\doibase 10.1371/journal.pcbi.1005659} {\bibfield  {journal} {\bibinfo  {journal} {PLOS Computational Biology}\ }\textbf {\bibinfo {volume} {13}},\ \bibinfo {pages} {1--17} (\bibinfo {year} {2017})}\BibitemShut {NoStop}%
\bibitem [{\citenamefont {Wang}\ \emph {et~al.}(2004)\citenamefont {Wang}, \citenamefont {Wolf}, \citenamefont {Caldwell}, \citenamefont {Kollman},\ and\ \citenamefont {Case}}]{Wang2004}%
  \BibitemOpen
  \bibfield  {author} {\bibinfo {author} {\bibfnamefont {J.}~\bibnamefont {Wang}}, \bibinfo {author} {\bibfnamefont {R.~M.}\ \bibnamefont {Wolf}}, \bibinfo {author} {\bibfnamefont {J.~W.}\ \bibnamefont {Caldwell}}, \bibinfo {author} {\bibfnamefont {P.~A.}\ \bibnamefont {Kollman}}, \ and\ \bibinfo {author} {\bibfnamefont {D.~A.}\ \bibnamefont {Case}},\ }\bibfield  {title} {\enquote {\bibinfo {title} {Development and testing of a general amber force field},}\ }\href {\doibase https://doi.org/10.1002/jcc.20035} {\bibfield  {journal} {\bibinfo  {journal} {J. Comp. Chem.}\ }\textbf {\bibinfo {volume} {25}},\ \bibinfo {pages} {1157--1174} (\bibinfo {year} {2004})}\BibitemShut {NoStop}%
\bibitem [{\citenamefont {Wang}, \citenamefont {Martinez},\ and\ \citenamefont {Pande}(2014)}]{wang2014building}%
  \BibitemOpen
  \bibfield  {author} {\bibinfo {author} {\bibfnamefont {L.-P.}\ \bibnamefont {Wang}}, \bibinfo {author} {\bibfnamefont {T.~J.}\ \bibnamefont {Martinez}}, \ and\ \bibinfo {author} {\bibfnamefont {V.~S.}\ \bibnamefont {Pande}},\ }\bibfield  {title} {\enquote {\bibinfo {title} {Building force fields: An automatic, systematic, and reproducible approach},}\ }\href@noop {} {\bibfield  {journal} {\bibinfo  {journal} {J. Phys. Chem. Lett.}\ }\textbf {\bibinfo {volume} {5}},\ \bibinfo {pages} {1885--1891} (\bibinfo {year} {2014})}\BibitemShut {NoStop}%
\bibitem [{\citenamefont {Izaguirre}, \citenamefont {Sweet},\ and\ \citenamefont {Pande}(2010)}]{Izaguirre2010}%
  \BibitemOpen
  \bibfield  {author} {\bibinfo {author} {\bibfnamefont {J.~A.}\ \bibnamefont {Izaguirre}}, \bibinfo {author} {\bibfnamefont {C.~R.}\ \bibnamefont {Sweet}}, \ and\ \bibinfo {author} {\bibfnamefont {V.~S.}\ \bibnamefont {Pande}},\ }\bibfield  {title} {\enquote {\bibinfo {title} {Multiscale dynamics of macromolecules using normal mode langevin},}\ }\href@noop {} {\bibfield  {journal} {\bibinfo  {journal} {Pac. Symp. Biocomput.}\ }\textbf {\bibinfo {volume} {15}},\ \bibinfo {pages} {240--251} (\bibinfo {year} {2010})}\BibitemShut {NoStop}%
\bibitem [{\citenamefont {Darden}, \citenamefont {York},\ and\ \citenamefont {Pedersen}(1993)}]{Darden1993}%
  \BibitemOpen
  \bibfield  {author} {\bibinfo {author} {\bibfnamefont {T.}~\bibnamefont {Darden}}, \bibinfo {author} {\bibfnamefont {D.}~\bibnamefont {York}}, \ and\ \bibinfo {author} {\bibfnamefont {L.}~\bibnamefont {Pedersen}},\ }\bibfield  {title} {\enquote {\bibinfo {title} {{Particle mesh Ewald: An Nlog(N) method for Ewald sums in large systems}},}\ }\href {\doibase 10.1063/1.464397} {\bibfield  {journal} {\bibinfo  {journal} {J. Chem. Phys.}\ }\textbf {\bibinfo {volume} {98}},\ \bibinfo {pages} {10089--10092} (\bibinfo {year} {1993})}\BibitemShut {NoStop}%
\bibitem [{\citenamefont {Floquet}\ \emph {et~al.}(2004)\citenamefont {Floquet}, \citenamefont {Héry-Huynh}, \citenamefont {Dauchez}, \citenamefont {Derreumaux}, \citenamefont {Tamburro},\ and\ \citenamefont {Alix}}]{Floquet2004}%
  \BibitemOpen
  \bibfield  {author} {\bibinfo {author} {\bibfnamefont {N.}~\bibnamefont {Floquet}}, \bibinfo {author} {\bibfnamefont {S.}~\bibnamefont {Héry-Huynh}}, \bibinfo {author} {\bibfnamefont {M.}~\bibnamefont {Dauchez}}, \bibinfo {author} {\bibfnamefont {P.}~\bibnamefont {Derreumaux}}, \bibinfo {author} {\bibfnamefont {A.~M.}\ \bibnamefont {Tamburro}}, \ and\ \bibinfo {author} {\bibfnamefont {A.~J.~P.}\ \bibnamefont {Alix}},\ }\bibfield  {title} {\enquote {\bibinfo {title} {Structural characterization of vgvapg, an elastin-derived peptide},}\ }\href {\doibase https://doi.org/10.1002/bip.20029} {\bibfield  {journal} {\bibinfo  {journal} {Peptide Science}\ }\textbf {\bibinfo {volume} {76}},\ \bibinfo {pages} {266--280} (\bibinfo {year} {2004})}\BibitemShut {NoStop}%
\bibitem [{\citenamefont {Donati}\ and\ \citenamefont {Keller}(2018)}]{Donati2018}%
  \BibitemOpen
  \bibfield  {author} {\bibinfo {author} {\bibfnamefont {L.}~\bibnamefont {Donati}}\ and\ \bibinfo {author} {\bibfnamefont {B.~G.}\ \bibnamefont {Keller}},\ }\bibfield  {title} {\enquote {\bibinfo {title} {Girsanov reweighting for metadynamics simulations},}\ }\href@noop {} {\bibfield  {journal} {\bibinfo  {journal} {J. Chem. Phys.}\ }\textbf {\bibinfo {volume} {149}},\ \bibinfo {pages} {072335} (\bibinfo {year} {2018})}\BibitemShut {NoStop}%
\bibitem [{\citenamefont {Giraldo-Barreto}\ \emph {et~al.}(2021)\citenamefont {Giraldo-Barreto}, \citenamefont {Ortiz}, \citenamefont {Thiede}, \citenamefont {Palacio-Rodríguez}, \citenamefont {Carpenter}, \citenamefont {Barnett},\ and\ \citenamefont {Cossio}}]{Barreto2021}%
  \BibitemOpen
  \bibfield  {author} {\bibinfo {author} {\bibfnamefont {J.}~\bibnamefont {Giraldo-Barreto}}, \bibinfo {author} {\bibfnamefont {S.}~\bibnamefont {Ortiz}}, \bibinfo {author} {\bibfnamefont {E.}~\bibnamefont {Thiede}}, \bibinfo {author} {\bibfnamefont {K.}~\bibnamefont {Palacio-Rodríguez}}, \bibinfo {author} {\bibfnamefont {B.}~\bibnamefont {Carpenter}}, \bibinfo {author} {\bibfnamefont {A.}~\bibnamefont {Barnett}}, \ and\ \bibinfo {author} {\bibfnamefont {P.}~\bibnamefont {Cossio}},\ }\bibfield  {title} {\enquote {\bibinfo {title} {A bayesian approach to extracting free-energy profiles from cryo-electron microscopy experiments},}\ }\href {\doibase 10.1038/s41598-021-92621-1} {\bibfield  {journal} {\bibinfo  {journal} {Scientific Reports}\ }\textbf {\bibinfo {volume} {11}} (\bibinfo {year} {2021}),\ 10.1038/s41598-021-92621-1}\BibitemShut {NoStop}%
\bibitem [{\citenamefont {Maier}\ \emph {et~al.}(2015)\citenamefont {Maier}, \citenamefont {Martinez}, \citenamefont {Kasavajhala}, \citenamefont {Wickstrom}, \citenamefont {Hauser},\ and\ \citenamefont {Simmerling}}]{Maier2015}%
  \BibitemOpen
  \bibfield  {author} {\bibinfo {author} {\bibfnamefont {J.~A.}\ \bibnamefont {Maier}}, \bibinfo {author} {\bibfnamefont {C.}~\bibnamefont {Martinez}}, \bibinfo {author} {\bibfnamefont {K.}~\bibnamefont {Kasavajhala}}, \bibinfo {author} {\bibfnamefont {L.}~\bibnamefont {Wickstrom}}, \bibinfo {author} {\bibfnamefont {K.~E.}\ \bibnamefont {Hauser}}, \ and\ \bibinfo {author} {\bibfnamefont {C.}~\bibnamefont {Simmerling}},\ }\bibfield  {title} {\enquote {\bibinfo {title} {ff14sb: Improving the accuracy of protein side chain and backbone parameters from ff99sb},}\ }\href@noop {} {\bibfield  {journal} {\bibinfo  {journal} {J. Chem. Theory Comput.}\ }\textbf {\bibinfo {volume} {11 (8)}},\ \bibinfo {pages} {3696--3713} (\bibinfo {year} {2015})}\BibitemShut {NoStop}%
\bibitem [{\citenamefont {Chiu}\ \emph {et~al.}(2005)\citenamefont {Chiu}, \citenamefont {Kubelka}, \citenamefont {Herbst-Irmer}, \citenamefont {Eaton}, \citenamefont {Hofrichter},\ and\ \citenamefont {Davies}}]{chiu2005high}%
  \BibitemOpen
  \bibfield  {author} {\bibinfo {author} {\bibfnamefont {T.~K.}\ \bibnamefont {Chiu}}, \bibinfo {author} {\bibfnamefont {J.}~\bibnamefont {Kubelka}}, \bibinfo {author} {\bibfnamefont {R.}~\bibnamefont {Herbst-Irmer}}, \bibinfo {author} {\bibfnamefont {W.~A.}\ \bibnamefont {Eaton}}, \bibinfo {author} {\bibfnamefont {J.}~\bibnamefont {Hofrichter}}, \ and\ \bibinfo {author} {\bibfnamefont {D.~R.}\ \bibnamefont {Davies}},\ }\bibfield  {title} {\enquote {\bibinfo {title} {High-resolution x-ray crystal structures of the villin headpiece subdomain, an ultrafast folding protein},}\ }\href@noop {} {\bibfield  {journal} {\bibinfo  {journal} {Proceedings of the National Academy of Sciences}\ }\textbf {\bibinfo {volume} {102}},\ \bibinfo {pages} {7517--7522} (\bibinfo {year} {2005})}\BibitemShut {NoStop}%
\bibitem [{\citenamefont {{Schrödinger, LLC}}(2015)}]{PyMOL}%
  \BibitemOpen
  \bibfield  {author} {\bibinfo {author} {\bibnamefont {{Schrödinger, LLC}}},\ }\href {https://pymol.org/2/} {\emph {\bibinfo {title} {{The PyMOL Molecular Graphics System, Version 2.0}}}},\ \bibinfo {organization} {Schrödinger, LLC} (\bibinfo {year} {2015}),\ \bibinfo {note} {accessed: 2024-10-08}\BibitemShut {NoStop}%
\bibitem [{\citenamefont {Prinz}\ \emph {et~al.}(2011)\citenamefont {Prinz}, \citenamefont {Wu}, \citenamefont {Sarich}, \citenamefont {Keller}, \citenamefont {Senne}, \citenamefont {Held}, \citenamefont {Chodera}, \citenamefont {Sch{\"{u}}tte},\ and\ \citenamefont {No{\'{e}}}}]{Prinz2011}%
  \BibitemOpen
  \bibfield  {author} {\bibinfo {author} {\bibfnamefont {J.-H.}\ \bibnamefont {Prinz}}, \bibinfo {author} {\bibfnamefont {H.}~\bibnamefont {Wu}}, \bibinfo {author} {\bibfnamefont {M.}~\bibnamefont {Sarich}}, \bibinfo {author} {\bibfnamefont {B.}~\bibnamefont {Keller}}, \bibinfo {author} {\bibfnamefont {M.}~\bibnamefont {Senne}}, \bibinfo {author} {\bibfnamefont {M.}~\bibnamefont {Held}}, \bibinfo {author} {\bibfnamefont {J.~D.}\ \bibnamefont {Chodera}}, \bibinfo {author} {\bibfnamefont {C.}~\bibnamefont {Sch{\"{u}}tte}}, \ and\ \bibinfo {author} {\bibfnamefont {F.}~\bibnamefont {No{\'{e}}}},\ }\bibfield  {title} {\enquote {\bibinfo {title} {{Markov models of molecular kinetics: generation and validation.}}}\ }\href@noop {} {\bibfield  {journal} {\bibinfo  {journal} {J. Chem. Phys.}\ }\textbf {\bibinfo {volume} {134}},\ \bibinfo {pages} {174105} (\bibinfo {year} {2011})}\BibitemShut {NoStop}%
\bibitem [{\citenamefont {Husic}\ and\ \citenamefont {Pande}(2018)}]{Pande2018}%
  \BibitemOpen
  \bibfield  {author} {\bibinfo {author} {\bibfnamefont {B.~E.}\ \bibnamefont {Husic}}\ and\ \bibinfo {author} {\bibfnamefont {V.~S.}\ \bibnamefont {Pande}},\ }\bibfield  {title} {\enquote {\bibinfo {title} {Markov state models: From an art to a science},}\ }\href {\doibase 10.1021/jacs.7b12191} {\bibfield  {journal} {\bibinfo  {journal} {Journal of the American Chemical Society}\ }\textbf {\bibinfo {volume} {140}},\ \bibinfo {pages} {2386--2396} (\bibinfo {year} {2018})}\BibitemShut {NoStop}%
\bibitem [{\citenamefont {Bowman}, \citenamefont {Pande},\ and\ \citenamefont {No{\'{e}}}(2014)}]{Bowman2014}%
  \BibitemOpen
  \bibinfo {editor} {\bibfnamefont {G.~R.}\ \bibnamefont {Bowman}}, \bibinfo {editor} {\bibfnamefont {V.~S.}\ \bibnamefont {Pande}}, \ and\ \bibinfo {editor} {\bibfnamefont {F.}~\bibnamefont {No{\'{e}}}},\ eds.,\ \href@noop {} {\emph {\bibinfo {title} {{An Introduction to Markov State Models and Their Application to Long Timescale Molecular Simulation}}}},\ Vol.\ \bibinfo {volume} {797 of Advances in Experimental Medicine and Biology}\ (\bibinfo  {publisher} {Springer},\ \bibinfo {address} {Heidelberg},\ \bibinfo {year} {2014})\BibitemShut {NoStop}%
\bibitem [{\citenamefont {Keller}, \citenamefont {Aleksic},\ and\ \citenamefont {Donati}(2018)}]{keller2018}%
  \BibitemOpen
  \bibfield  {author} {\bibinfo {author} {\bibfnamefont {B.~G.}\ \bibnamefont {Keller}}, \bibinfo {author} {\bibfnamefont {S.}~\bibnamefont {Aleksic}}, \ and\ \bibinfo {author} {\bibfnamefont {L.}~\bibnamefont {Donati}},\ }\bibfield  {title} {\enquote {\bibinfo {title} {Markov state models in drug design},}\ }in\ \href@noop {} {\emph {\bibinfo {booktitle} {Biomolecular Simulations in Structure-based Drug Discovery}}},\ \bibinfo {editor} {edited by\ \bibinfo {editor} {\bibfnamefont {F.~L.}\ \bibnamefont {Gervasio}}}\ (\bibinfo  {publisher} {Wiley-Interscience},\ \bibinfo {address} {Weinheim},\ \bibinfo {year} {2018})\ p.~\bibinfo {pages} {67}\BibitemShut {NoStop}%
\bibitem [{\citenamefont {Harada}\ and\ \citenamefont {Kitao}(2012)}]{Harada2012}%
  \BibitemOpen
  \bibfield  {author} {\bibinfo {author} {\bibfnamefont {R.}~\bibnamefont {Harada}}\ and\ \bibinfo {author} {\bibfnamefont {A.}~\bibnamefont {Kitao}},\ }\bibfield  {title} {\enquote {\bibinfo {title} {The fast-folding mechanism of villin headpiece subdomain studied by multiscale distributed computing},}\ }\href {\doibase 10.1021/ct200363h} {\bibfield  {journal} {\bibinfo  {journal} {Journal of Chemical Theory and Computation}\ }\textbf {\bibinfo {volume} {8}},\ \bibinfo {pages} {290--299} (\bibinfo {year} {2012})}\BibitemShut {NoStop}%
\bibitem [{\citenamefont {Wang}\ \emph {et~al.}(2019)\citenamefont {Wang}, \citenamefont {Tao}, \citenamefont {Wang},\ and\ \citenamefont {Xiao}}]{Wang2019}%
  \BibitemOpen
  \bibfield  {author} {\bibinfo {author} {\bibfnamefont {E.}~\bibnamefont {Wang}}, \bibinfo {author} {\bibfnamefont {P.}~\bibnamefont {Tao}}, \bibinfo {author} {\bibfnamefont {J.}~\bibnamefont {Wang}}, \ and\ \bibinfo {author} {\bibfnamefont {Y.}~\bibnamefont {Xiao}},\ }\bibfield  {title} {\enquote {\bibinfo {title} {A novel folding pathway of the villin headpiece subdomain hp35},}\ }\href {\doibase 10.1039/C9CP01703H} {\bibfield  {journal} {\bibinfo  {journal} {Physical Chemistry Chemical Physics}\ }\textbf {\bibinfo {volume} {21}} (\bibinfo {year} {2019}),\ 10.1039/C9CP01703H}\BibitemShut {NoStop}%
\end{thebibliography}%
